\documentclass[12pt,a4paper,notitlepage]{article}

\usepackage{graphicx,psfrag} 
\usepackage{enumerate}
\usepackage{subfig}
\usepackage{natbib}
\usepackage[colorlinks=true, allcolors=blue]{hyperref}
\usepackage{secdot}
\usepackage{amsmath,amssymb}
\usepackage{todonotes}
\usepackage{lineno}
\usepackage{comment}
\usepackage{xargs}                      
\usepackage{apalike}
\usepackage{float}

\usepackage{setspace}
\usepackage{secdot}

\newcommandx{\rk}[2][1=]{\todo[inline,backgroundcolor=blue!25,bordercolor=green,#1]{Ruth: #2}}
\newcommandx{\bs}[2][1=]{\todo[inline,backgroundcolor=cyan!35,bordercolor=black,#1]{Blanca: #2}}
\newcommandx{\ve}[2][1=]{\todo[inline,backgroundcolor=blue!15!white,bordercolor=blue,#1]{Victor: #2}}

\newcommand{\mcJ}{\mathcal{J}}

\newcommand{\bfepsilon}{\hbox{\boldmath$\epsilon$}}

\newcommand{\bftheta}{\hbox{\boldmath$\theta$}}

\newcommand{\bfphi}{\hbox{\boldmath$\phi$}}

\newcommand{\bfomega}{\hbox{\boldmath$\omega$}}
\newcommand{\bfsomega}{\hbox{\boldmath{\scriptsize{$\omega$}}}}

\newcommand{\bfp}{\hbox{\boldmath$p$}}

\newcommand{\bfx}{\hbox{\boldmath$x$}}

\title{Large Data and (Not Even Very) Complex Ecological Models: When Worlds Collide}

\author{Ruth King$^1$, Blanca Sarzo$^{1,2}$, and V\'ictor Elvira$^1$}
\date{%
$^1$School of Mathematics and Maxwell Institute for Mathematical Sciences, 
University of Edinburgh, Edinburgh, UK \\
$^2$Department of Microbiology and Ecology, University of Valencia, Valencia, Spain}

\begin{document}

\maketitle

\begin{abstract}
We consider the challenges that arise when fitting complex ecological models to ``large'' data sets. In particular, we focus on random effect models which are commonly used to describe individual heterogeneity, often present in ecological populations under study. In general, these models lead to a likelihood that is expressible only as an analytically intractable integral. Common techniques for fitting such models to data include, for example, the use of numerical approximations for the integral, or a Bayesian data augmentation approach. However, as the size of the data set increases (i.e. the number of individuals increases), these computational tools may become computationally infeasible. We present an efficient Bayesian model-fitting approach, whereby we initially sample from the posterior distribution of a smaller subsample of the data, before correcting this sample to obtain estimates of the posterior distribution of the full dataset, using an importance sampling approach. We consider several practical issues, including the subsampling mechanism, computational efficiencies (including the ability to parallelise the algorithm) and combining subsampling estimates using multiple subsampled datasets. We demonstrate the approach in relation to individual heterogeneity capture-recapture models. We initially demonstrate the feasibility of the approach via simulated data before considering a challenging real dataset of approximately 30,000 guillemots, and obtain posterior estimates in substantially reduced computational time. 
\end{abstract}

sampling; individual heterogeneity; intractable likelihood; random effects.

\section{Introduction}

The use of continuous random effect models within statistical ecology applications is becoming increasingly common, particularly where individual and/or temporal heterogeneity can be substantial \citep{gimcg17}. However, the introduction of such random effects often leads to a likelihood that is expressible only in the form of an analytically intractable integral. We focus on the inclusion of individual heterogeneity within the Cormack-Jolly-Seber (CJS) model for capture-recapture data, where the survival probabilities are the primary parameters of interest, and on which we wish to specify individual heterogeneity.

Traditionally, many different approaches have been applied to obtain estimates of the model parameters when the likelihood is analytically intractable. For example, within a classical framework, numerical integration schemes have been applied such as Gaussian-Hermite quadrature for low dimensional problems \citep{coua99,gimenez10}; Laplace approximations \citep{herkk21}; Monte Carlo-type estimates for higher dimensional integrals \citep{dev02,dev04}; and the reduction to finite mixture models \citep{ple00,plepn03}. Alternatively, within a Bayesian approach data augmentation (or complete-data likelihood approach) have been applied \citep{kinb08,royle2008modeling,kinmkb16}.

Large scale capture-recapture-type studies are becoming increasingly common where several thousands of individuals may be ringed/tagged each year. This is particularly true for bird studies. For example, \cite{Hestbeck91} consider data relating to nearly 30,000 Canada Geese; while \cite{Francis09} has data from approximately 20,000 Tawny Owls. 
However, many traditional model-fitting approaches for heterogeneity models do not scale when the dataset becomes ``large'', in terms of the number of individuals in the study; and/or when the likelihood increases in complexity due to the given the model structure. 

More generally within the wider statistical literature, for large dataset two approaches are often used: (i) divide-and conquer that partitions the data into multiple datasets, analysing each independently and recombining; and (ii) using a subsample of the data to estimate the full posterior. See \cite{bar17} for further discussion. Our approach focuses on the latter idea. In particular, we propose an algorithm that initially analyses a smaller subsample of the data using a Markov chain Monte Carlo (MCMC) sampler \citep{mcmc}, and then corrects the sampled parameter values such that we obtain an estimate of the posterior distribution of the full dataset of interest. The subsampled data are such that a Bayesian data augmentation approach can be applied within standard black-box software. The realisations of the Markov chain are then reweighted via an importance sampling algorithm \cite[for a review of importance sampling, see for example,][]{tokk10,elvm21} to obtain an estimate of the posterior distribution for the full dataset. Multiple sets of subsampled data can be taken and analysed in parallel, independently of each other, and subsequently combined to decrease the Monte Carlo error of the posterior estimates. 

In Section \ref{model-case}, we describe the CJS model and motivating case study relating to common guillemots (\textit{Uria aalge}). In Section \ref{method}, we describe the model-fitting algorithm of subsampling the data, and subsequently correcting the output via  importance sampling, before discussing associated practical implementation issues in Section \ref{practical}. We apply the approach to a simulated dataset in Section \ref{sim} and the case study in Section \ref{data}, for which the traditional Bayesian data augmentation technique is computationally infeasible. We conclude in Section \ref{disc}.

\vspace{-3 truemm}

\section{Model description and case study}\label{model-case}

We first introduce the CJS model before presenting the common guillemot case study. 

\vspace{-2 truemm}

\subsection{Cormack-Jolly-Seber model}

We consider capture-recapture studies, where data are collected over a series of discrete capture occasions, $t=1,\dots,T$. At each occasion, all observed individuals are recorded. The first time an individual is observed, an associated unique identifier is recorded (e.g. natural skin/fur markings) or applied (e.g. a physical ring/tag attached). The capture-recapture data are the associated capture histories of each individual observed within the study, $i=1,\dots,I$, indicating whether the given individual was observed or not at each capture occasion. Mathematically, for $i=1,\dots,I$ and $t=1,\dots,T$, we let,
\begin{linenomath*}
\begin{align}  
x_{it} = \left\{ \begin{array}{ll}
0     & \mbox{ if individual $i$ is not observed at time $t$;}  \\
1     & \mbox{ if individual $i$ is observed at time $t$.} 
\end{array} \right.
\end{align}
\end{linenomath*}
We let $f_i$ and $l_i$ denote the first and last time individual $i=1,\dots,I$ is observed in the study. The capture history for individual $i=1,\dots,I$ is denoted $\bfx_i = \{x_{it}:t=1,\dots,T\}$; with the full dataset, $\bfx = \{\bfx_i:i=1,\dots,I\}$. 
We consider only live recaptures but the approach is immediately extendable to include dead recoveries. 
The CJS model 
conditions on initial capture and is defined in terms of (\textit{apparent}) survival and recapture probabilities. Mathematically for $i=1,\dots,I$ and $t=1,\dots,T-1$, we define:
\begin{description}
\item[ ] ~ $\phi_{it} = \mathbb{P}(\mbox{individual $i$ is alive at time $t+1$ $|$ alive at time $t$})$;
\item[ ] ~ $p_{it+1} = \mathbb{P}(\mbox{individual $i$ is observed at time $t+1$ $|$ alive at time $t+1$})$.
\end{description}
We let $\bfphi = \{\phi_{it}: i=1,\dots,I; t=1,\dots,T-1\}$ and $\bfp = \{p_{it}:i=1,\dots,I; t=2,\dots,T\}$.
More generally, the state of ``alive'' corresponds to being available for capture, so that $\phi_{it}$ corresponds to \textit{apparent} survival with emigration and survival confounded. For simplicity, we refer to $\phi_{it}$ as survival. 
The corresponding likelihood can be expressed in the form, 
\begin{linenomath*}
\begin{equation*}
f(\bfx | \bfphi, \bfp) = \prod_{i=1}^I f(\bfx_i | \bfphi, \bfp).
\end{equation*}
\end{linenomath*}
The term $f(\bfx_i | \bfphi, \bfp)$ denotes the probability of the capture history of individual $i$ given by,
\begin{linenomath*} 
\begin{equation}\label{eq1}
    f(\bfx_i | \bfphi, \bfp) = \left[ \prod_{t=f_i}^{l_i-1}  \phi_{it} \: p_{it+1}^{x_{it+1}} (1-p_{it+1})^{(1-x_{it+1})} \right] \times \chi_{il_i},
\end{equation}
\end{linenomath*}
where $\prod_{t=f_i}^{f_i-1} \equiv 1$; and $\chi_{it}$ denotes the probability individual $i$ is not observed after time $t$, given they are alive at $t$. This probability is most often described via the recursion,

\vspace{-8 truemm}
\begin{linenomath*}
\[
\chi_{it} = 1 - \phi_{it}(1-(1-p_{it+1})\chi_{it+1}), \qquad \mbox{ with } \quad \chi_{iT} = 1.
\]
\end{linenomath*}
See \citealp{king2010bayesian, king14, McCrea, seber19} for further details and a comprehensive review of capture-recapture-type models. 

\subsection{Individual random effect models}

\vspace{-1 truemm}
We consider the case where the random effects are specified on the survival probabilities:
\vspace{-1 truemm}
\begin{linenomath*}
\[
\mbox{logit } \phi_{it} = \alpha + \epsilon_i, \qquad \mbox{ where } \epsilon_i \sim N(0,\sigma^2),
\]
\end{linenomath*}
for $t=1,\dots,T-1$ and $i=1, \ldots, I$. The model parameters are denoted $\bftheta = \{\alpha,\bfp,\sigma^2\}$, with the random effects, $\bfepsilon = \{\epsilon_i:i=1,\dots,I\}$  integrated out in the observed data likelihood:
\begin{linenomath*}
\begin{equation}\label{lik1}
f(\bfx | \alpha, \bfp, \sigma^2) = \prod_{i=1}^I \int_{\epsilon_i} f(\bfx_i | \alpha, \bfp, \epsilon_i) f(\epsilon_i | \sigma^2) d \epsilon_i,
\end{equation}
\end{linenomath*}
where $f(\bfx_i | \alpha, \bfp, \epsilon_i)$ is as in Equation (\ref{eq1}); and $f(\epsilon_i|\sigma^2)$ denotes the random effect density, which in our case study, we assume to be Gaussian. The approach immediately generalises to random effects on all model parameters and mixed-effects type model, allowing for additional temporal or covariate effects and non-Gaussian random effect distributions.

\vspace{-2 truemm}

\subsection{Case study: guillemots}\label{case}

\vspace{-1 truemm}
We consider capture-recapture data of guillemots on the island of Stora Karls\"{o} (Sweden). This is the largest guillemot colony in the Baltic Sea with a recorded breeding population of 15,700 pairs in 2014, corresponding to $\approx$ 2/3 of the Baltic Sea population \citep{olsson2017}. We consider data from 2006-2016 (i.e. $T=11$), with a total of $I=28,930$ birds ringed. Recaptures were via resightings during the breeding season (May to July) using long-sighted telescopes. 
For further details see, for example, \cite{Sarzo19}. Previous work by \cite{Sarzo21} suggested the presence of individual heterogeneity within the survival process, but due to the computational challenges was not investigated further. 

\vspace{-4 truemm}

\section{Method}\label{method}

The observed data likelihood in Equation \eqref{lik1} is analytically intractable. 
To fit such models numerical integration techniques may be used to estimate the integral over the individual heterogeneity terms  \citep[e.g.][]{gimenez10, coua99} or a Bayesian data augmentation approach applied \citep{royle2008modeling,king2010bayesian}. However, as the number or dimension of the random effects increases and/or the model increases in complexity, these approaches become computationally more challenging. We propose a Bayesian model-fitting approach that is scalable to large datasets and more complex models. 
The idea involves initially fitting the random effects model using a subsample of the data, and then correcting the sampled values using an associated importance weight. In this way, it is possible to approximate posterior summary statistics with consistent importance sampling estimators.


\vspace{-1 truemm}

\subsection{Algorithm} \label{IS}
The algorithm involves initially subsampling the data, and forming the posterior distribution of the model parameters, given the subsampled data, hereafter referred to as the \emph{subposterior}. In our case, the subsampling is at the individual capture history level. The subsampled data are designed such that it is computationally feasible, using a standard Bayesian data augmentation technique, to obtain a set of sampled parameter values from the subposterior (see for example, \cite{royle2008modeling,king2010bayesian}). In particular, we implement the Markov chain Monte Carlo (MCMC) sampling approach described by \cite{gimenez2007state}, additionally imputing the live/dead (or more accurately available/unavailable for capture) state of each individual following initial capture. We then correct this set of sampled parameter values by taking into account the remaining (unsampled) data via importance sampling, i.e. by assigning each sampled value with an importance weight to estimate the posterior distribution of the full data. The algorithm can be summarised as follows:
\begin{description}
\item \textbf{Step 1}: Draw a (random) subsample of the data by sampling without replacement a set of individuals from the set of observed individuals. 
\item \textbf{Step 2}: Using the set of subsampled individuals, implement a standard Bayesian MCMC data augmentation approach to obtain a set of sampled values from the given subposterior.
\item \textbf{Step 3}: Apply an importance sampling algorithm to correct the sampled values from the subposterior (by assigning an importance weight to each of them)  to obtain estimates of the posterior distribution of the full dataset. 
\end{description}
Steps 1-3 provide an estimate of the posterior distribution of the parameters. However, the steps can be repeated multiple times to obtain multiple estimates of the posterior distribution. Thus we advocate for an additional step to improve the estimation procedure:
\begin{description}
\item \textbf{Step 4}: Repeat Steps 1-3 a total of $M$ times and combine the posterior estimates of the parameters to obtain an improved estimate of the full posterior distribution.
\end{description}

Steps 1-3 can be undertaken in parallel across each of the subsamples $m=1,\dots,M$ as they are independent of each other. Thus, these steps are embarrassingly parallisable so that using multiple cores will significantly improve the computational efficiency of the algorithm. Although Step 4 is not strictly necessary, as each posterior obtained for a given subsample is an estimate of the posterior distribution of the parameters given the full data set, combining multiple posterior estimates improves the robustness and precision of the estimated posterior distribution. We now describe in further detail each individual steps.

\vspace{-3 truemm}

\subsubsection*{Step 1 - Subsampling the data:} ~ 
Recall that the dataset is denoted by $\bfx = \{\bfx_i: i=1,\dots,I\}$. We define a subsampled dataset by $\bfx^1 = \{\bfx_j: j \in \mcJ\}$, where $\mcJ \subset \{1,\dots,I\}$  denotes the elements of the data that are contained in the given subsample. The associated individual random effects are denoted by $\bfepsilon^1 = \{\epsilon_j: j \in \mcJ\}$. Further, we let $\mcJ^c = \{1,\dots,I\}\backslash \mcJ$ denote the complement of $\mcJ$ corresponding to the set of non-subsampled individuals, with associated capture histories $\bfx^2$ (so that $\bfx^2 = \bfx \backslash \bfx^1$). We refer to $\bfx^2$ as the \textit{remaining} data.

The simplest sampling scheme is to sample uniformly over individuals. However, this scheme can lead to poor precision due to large sampling variability. Alternatively, stratified sampling may be applied, for suitably defined strata (such as via cohort or other characteristics) to reduce this variability. We discuss different subsampling schemes in Section \ref{sample}. 

\vspace{-3 truemm}

\subsubsection*{Step 2 - Sampling from the subposterior:} ~
For a given subsample of the data, $\bfx^1$, we form the corresponding subposterior of the model parameters given by,
\begin{linenomath*}
\[
\pi^1(\bftheta | \bfx^1) \propto f(\bfx^1 | \bftheta) p(\bftheta).
\]
\end{linenomath*}
The likelihood $f(\bfx^1 | \bftheta)$ is analytically intractable. We implement a data augmentation scheme, with auxiliary variables $\bfepsilon^1$, to obtain a set of sampled values from $\pi^1(\bftheta | \bfx^1)$. Mathematically, we form the joint subposterior distribution of the parameters and auxiliary variables:
\begin{linenomath*}
\[
\pi^1(\bftheta, \bfepsilon^1 | \bfx^1) \propto f(\bfx^1 | \bftheta, \bfepsilon^1) f(\bfepsilon^1|\bftheta) p(\bftheta).
\]
\end{linenomath*}
We assume that we can use a standard MCMC algorithm to obtain a set of $N$ sampled values $\{\bftheta_k, \bfepsilon^1_k: k=1,\dots,K\}$ following a suitable burn-in period \citep{gelman2014bayesian,robert2018accelerating,van21}. For example, black-box software, such as BUGS \citep{lunn2000}, JAGS \citep{plummer2003jags}, NIMBLE 
\citep{nimble} or Stan \citep{carpenter}, may be used to obtain posterior samples from $\pi^1(\bftheta, \bfepsilon^1 | \bfx^1)$. For specific code for capture-recapture models, see for example, \cite{gimenez09,king2010bayesian,kery2011bayesian}. 
Note that there is control over the size of the subsample, so we can ensure a feasible computational time for obtaining the set of subposterior sampled values. We discuss the practical considerations regarding subsample size in Section \ref{size}. 

The sampled values can be used to obtain estimates of the marginal subposterior summary statistics of interest, $\pi^1(\bftheta|\bfx^1)$. 
However, we are interested in the posterior distribution of the full data, $\pi(\bftheta|\bfx)$, which we refer to as the \emph{full posterior distribution}. 
We would expect that the subposterior would be similar to the full posterior distribution but not identical. More precisely, we would expect the subposterior density to be wider (or more overdispersed) compared to the  posterior distribution given the full dataset, due to a reduction of information in the subsample. 
In order to account for the full dataset, we apply a correction to the parameter values simulated from the subposterior using an importance sampling algorithm, which permits us to obtain estimates of the posterior distribution of the full dataset, $\bfx$. 

\vspace{-3 truemm}

\subsubsection*{Step 3 - Importance sampling:} ~ 
We implement an importance sampling step on the sampled parameters and random effect values, $(\bftheta_1, \bfepsilon^1_1), \dots, (\bftheta_K, \bfepsilon^1_K)$ where the corresponding proposal distribution is the subposterior, $\pi^1(\bftheta,\bfepsilon^1|\bfx^1)$, with target distribution,  $\pi(\bftheta,\bfepsilon^1|\bfx)$. 
For $k=1,\dots,K$, the corresponding importance sampling weight, $\{w_k\}_{k=1}^K$, is given by,
\begin{linenomath*}
\begin{eqnarray*}
w_k & = & \frac{\pi(\bftheta_k, \bfepsilon^1_k | \bfx)}{\pi^1(\bftheta_k, \bfepsilon^1_k | \bfx^1)} \propto \frac{f(\bfx | \bftheta_k, \bfepsilon^1_k) p(\bfepsilon^1_k | \bftheta_k) p(\bftheta_k)}{f(\bfx^1 | \bftheta_k, \bfepsilon^1_k) p(\bfepsilon^1_k | \bftheta_k) p(\bftheta_k)}\\
& = & f(\bfx^2 | \bftheta_k),
\end{eqnarray*}
\end{linenomath*}
where $f(\bfx^2 | \bftheta_k) = \prod_{i \in \mcJ^c} f(\bfx_i | \bftheta_k)$. In other words, the associated importance weight, $w_k$, is the observed data likelihood for $\bfx^2$ evaluated at $\bftheta_k$. However, this weight is again analytically intractable. We extend the importance sampling approach and replace the likelihood expression with an estimate of this function denoted $\widehat{f}(\bfx^2 | \bftheta_k)$, and estimate the weight as,
\begin{linenomath*}
\[
\widehat{w}_k \propto \widehat{f}(\bfx^2| \bftheta_k).
\]
\end{linenomath*}

\noindent \cite{tran16} show that if the estimate is unbiased i.e. $\mathbb{E}(\widehat{f}(\bfx^2| \bftheta_k)) = f(\bfx^2 | \bftheta_k)$ the corresponding importance sampling estimate converges almost surely to the distribution of interest (and termed this approach IS$^2$). The result is akin to that of \cite{Andrieu_2009, Andrieu_2010} for particle MCMC, where replacing the likelihood with an unbiased estimate within an MCMC algorithm leads to the desired posterior distribution. 

We propose a Monte Carlo (MC) approach to obtain $\widehat{w}_k$. 
In the simplest case, for each $i \in \mcJ^c$ 
we simulate $N$ values of the random effects, $\bfepsilon_i = \{\epsilon_i(1),\dots,\epsilon_i(N)\}$ such that $\epsilon_i(j) \sim N(0,\sigma^2_k)$ for $j=1,\dots,N$. The unnormalised importance sampling weight is estimated as,

\begin{linenomath*}
\begin{equation}\label{eq:mc1}
\widehat{w}_k^* = \prod_{i \in \mcJ^c} \left[\frac{1}{N} \sum_{j=1}^N  f(\bfx_i| \bftheta_k, \epsilon_i(j)) \right], 
\end{equation}
\end{linenomath*}
where $f(\bfx_i| \bftheta_k, \epsilon_i(j))$ denotes the closed form conditional likelihood contribution for capture history $\bfx_i$, given the model parameters, $\bftheta_k$ and associated individual random effect, $\epsilon_i(j)$. 

We subsequently estimate the normalised sampling weights $\{\widehat{w}_k: k=1,\dots,K\}$ using,
\begin{linenomath*}
\begin{equation}
\widehat{w}_k = \frac{\widehat{w}^*_k}{\sum_{j=1}^K \widehat{w}^*_j},
\label{eq_snis}
\end{equation}
\end{linenomath*}
through the self-normalized importance sampling (SNIS) estimator \citep{elvm21}. These importance sampling weights can be used to obtain summary statistics/distributions of interest. For example, to obtain the posterior mean of some parameter, $\theta_1$, we use,
\begin{linenomath*}
\[
\mathbb{E}_{\pi}(\theta_1) = \sum_{k=1}^N \widehat{w}_k \; \theta_1.
\]
\end{linenomath*}
A sampling importance resampling (SIR) approach can be used to obtain sample parameter values  which can be used to obtain density estimates and/or 95\% credible intervals (CIs).

The MC estimate of the likelihood may become computationally expensive as $N$ and $I$ increase. The MC estimates are required for each posterior subsample, $m=1,\dots,M$, although these computations are parallelisable across subsamples, $m=1,\dots,M$ and individuals $i \in \mcJ^c$. We discuss further computational considerations in Section \ref{practical} and suggest approaches to decrease the computational component, including a stratified MC estimate; two-step algorithm and approximate (biased but consistent) weight estimate. 

\subsubsection*{Step 4 - Combined posterior estimate:} ~
Steps 1-3 are embarrassingly parallelisable over $m=1,\dots,M$; each subsampled dataset is independently drawn and separate MCMC algorithms applied. (Step 3 is also parallelisable over MC samples). Thus for no extra computational cost we can obtain multiple estimates of the full posterior distribution (at least up to the number of processors available). 
These posterior distributions can be combined to obtain a more reliable and robust estimate of the full posterior.
The combined estimate of the posterior distribution of the full data is defined to be a (weighted) average of the $M$ subsample posterior distributions. For example, to obtain the posterior mean of the parameter, $\theta_1$ we use,

\vspace{-7 truemm}

\begin{linenomath*}
\begin{equation}
\mathbb{E}_{\pi}(\theta_1) = \sum_{m=1}^M z_m \mathbb{E}_{\pi(m)}(\theta_1),     
\label{eq_combination}
\end{equation}
\end{linenomath*}
where $\mathbb{E}_{\pi(m)}(\theta_1)$ denotes the posterior mean of $\theta_1$ given the full dataset estimated using subsampled data $m=1,\dots,M$; and $z_1,\dots,z_M$ are corresponding weights such that $\sum_{m=1}^M z_m = 1$ and $0 \le z_m \le 1$. We discuss different possible weights in Section \ref{combine}.

\vspace{-2 truemm}

\section{Practical considerations}\label{practical}

We now discuss some practical considerations relating to the proposed algorithm.

\vspace{-2 truemm}

\subsection{Subsample size}\label{size}

A decision within the algorithm relates to the proportion of the data to subsample (i.e. $|\bfx^1|$). The larger the subsample, the closer the subposterior should be to the full posterior, so that the importance sampling algorithm increases in efficiency; however also the larger the computational cost in sampling from the subposterior. This computational cost is in terms of (1) time per each iteration (due to the number of auxiliary variables and cost to evaluate the likelihood function), and (2) length of MCMC simulations required since poorer mixing is often observed due to increased correlation between the  parameters (notably for the random effects, $\bfepsilon^1$, and $\sigma^2$).
Alternatively, smaller subsamples provide subposteriors for which it is (relatively) computationally fast to obtain a sample from but where the following importance sampling algorithm may suffer from increased particle depletion due to differences between the subposterior and full posterior (see,  \citealp{elvm21} for further discussion). Further, in this case, there is an increased computational cost in the calculation of the importance sampling weight, as this is a function of the remaining data, though this is minimised when using an alternative (biased) weight calculation making use of repeated histories (see Section \ref{MC}, consideration (iii)). In practice, the proportion of the data to sample will be dependent on the computational resources available, with the general advice to take as large a sample as possible that is computationally reasonable. 
For both the simulation study and case study, a subsample size of 20\% appeared to be a good trade-off with similar subposterior to full posterior for a relatively low computational cost. 

\vspace{-2 truemm}

\subsection{Sampling schemes}\label{sample}

We focus on stochastic schemes to subsample datasets. Ideally, the subposterior should be as similar as possible to the full posterior, to maximise the efficiency of the importance sampling approach. Random subsampling, selecting each capture history with equal probability, ignores any structure within the data, and thus typically leads to relatively non-similar distributions and poor performance (this is easily seen via simulation). Thus we consider a stratified sampling approach, where we initially stratify the individual capture histories, and then perform proportional random sampling within each strata. This approach is designed to replicate data structures in the subsample that are present within the full dataset. For example, strata may be defined via observable covariate information (such as age/gender); cohort (i.e. year of first capture); unique capture histories; or capture histories with defined characteristics, such as the number of times observed alive; or initial and final capture times. In practice, it may also be desirable to pool several strata when frequency sizes are small. An `optimal' scheme will typically depend on the dependence of the model parameters, as for standard sampling techniques \citep{hanmn19}, and model being fitted to the data. For example, if the model parameters are assumed to be age dependent, then this suggests that including age within the subsampling stratification may be useful. 

\vspace{-1 truemm}

\subsection{Estimation of importance sampling weights}\label{MC}

We consider a MC approach for the estimation of the importance sampling weight, $w_k$. We focus on three approaches in relation to computational efficiency: (i) stratified MC; (ii) 2-step MC; and (iii) repeated histories. We discuss each in turn.

\vspace{-1 truemm}

\subsubsection*{(i) Stratified MC approach:} ~ For increased computational efficiency, we apply a stratified MC approach \citep[Chapter 8]{mcbook}. In particular, we partition $\mathbb{R}$ into $N$ strata, separated by the $N-1$ quantiles of the given $N(0,\sigma^2)$ distribution, and simulate a single particle in each strata. 
This leads to strata of varying length but, by definition, have equal probability. 
Consequently, the estimate for the unnormalised weight is as in Equation (\ref{eq:mc1}), due to the equal probability of each stratum, but reduces the associated variability of the estimate.

\vspace{-1 truemm}

\subsubsection*{(ii) Two-step MC approach:} ~ Despite attempts to minimise the difference between the subposterior and the full posterior, we will typically observe parameter values with negligible weight (with this ``particle depletion'' generally increasing with the number of parameters). To increase computational efficiency, we propose a two-step approach. In the first step, we use a coarse (stratified) MC approach to obtain an estimate of the (unnormalised) weight. In the second step, we retain only the top ranked sampled values (for example, the the top 10-20\%) or those with non-negligible weight and calculate a more accurate estimate of their associated weights using a much finer MC estimate. 
In practice, obtaining a fast and reliable ``ball-park'' value in the first step is generally straightforward. For example, for the real data application, using as few as $N=25$ MC values where we simulate a single value within each 4\% quantile range of the random effect distribution (or even use the mid-point of the quantile ranges) led to stable estimates in terms of the ranking of the sampled values to be retained for obtaining a more accurate MC estimate of the weight (the top 10\% were retained for Step 2). We note that the approach works when the variability within the MC estimates for given sampled parameter values is smaller than the variability of the weights across parameter values. 

\subsubsection*{(iii) Repeated histories:} ~ The weight in Equation (\ref{eq:mc1}) is a product over the number of individuals in $\bfx^2$, and thus scales linearly with the number of histories. However many individuals will have the same capture history, and hence marginal likelihood contribution. In other words, for individuals $i$ and $j$ that have the same history, $f(\bfx_i|\bftheta) = f(\bfx_j|\bftheta)$. In the MC scheme described above we obtain an estimate of the marginal likelihood for each individual, independently, leading to an unbiased estimate of the marginal likelihood, $f(\bfx^2|\bftheta)$. However, we can consider an alternative (biased) estimate of the weight that is computationally faster by only estimating the marginal likelihood for \textit{unique} histories. 

Let $\mcJ^c_U$ denote the set of unique capture histories in $\bfx^2$ and $n(\bfomega)$ the number of individuals in $\bfx^2$ with capture history $\bfomega \in \mcJ^c_U$. For each history $\bfomega \in \mcJ^c_U$ and sampled parameter value $\bftheta_k$, for $k=1,\dots,K$, we simulate $N$ values of the random effects $\bfepsilon_{\bfsomega} = \{\epsilon_{\bfsomega}(1),\dots,\epsilon_{\bfsomega}(N)\}$, such that $\epsilon^2_{\bfsomega}(i) \sim N(0,\sigma^2_k)$. 
We estimate the importance sampling weight using,
\begin{linenomath*}
\begin{equation*}
\widehat{w}^*_k = \prod_{\bfomega \in \mcJ^c_U} \left[\frac{1}{N} \sum_{j=1}^N f(\bfomega|\bftheta_k, \epsilon_{\bfsomega}(j)) \right]^{n(\bfsomega)},
\end{equation*}
\end{linenomath*}
where $f(\bfomega|\bftheta_k, \epsilon_{\bfsomega}(j))$ denotes the conditional likelihood contribution for capture history $\bfomega$, given $\bftheta_k$ and $\epsilon_{\bfsomega}(j)$. This estimate (though biased) is a consistent estimator of the unnormalised weight. We note that the number of unique capture histories is, in general, significantly smaller than the number of individuals observed, and hence scales significantly slower as the number of individuals increases. More precisely, the maximum number of unique capture histories is $2^T$ and hence limited by the number of capture occasions, and in most cases not all histories will be observed within the dataset. Thus, this estimate is significantly faster computationally in general, at the expense of the property of unbiasedness for finite sample size. Further, we note that given the substantially reduced required number of MC estimates at the capture history level, we can use a significantly larger value for $N$. In practice, for the case study in Section \ref{data}, where the number of individuals observed with the same capture history is of the order of 1000s, we use a hybrid approach. In this approach we essentially specify the data using multiple replicates of the same capture history, such that the number of individuals with each of these (repeated) capture histories is limited to be at most some specified maximum value (for the case study a value of 200 was used). Within the MC estimate of the unnormalised weight we then consider each of these histories as unique. This hybrid approach led to improved convergence of the MC estimate of the weight. 

\subsection{Combining multiple importance sampling estimates}\label{combine}

The importance sampling algorithm is naively parallelisable for the subsampled datasets. Thus, given sufficient computer cores, we can obtain multiple posterior estimates at no additional computational cost. Further, the estimates across different subsampled data can be combined to obtain an improved estimate of the full posterior, as in Equation (\ref{eq_combination}). The function is a linear combination of the posterior estimates for each subsampled dataset, 
for any set of positive weights that sum to unity. For example, in the simplest case, $z_m = \frac{1}{M}$, for $m=1,\dots,M$. However, this implicitly assumes that all the subsampled posterior estimates are equally informative, which in general will not be the case. To address this, we may consider, for example, setting $z_m$ as proportional to the inverse of the variance of the weights \citep{Douc07,luengo2018efficient} or effective sample size (or unique number of non-negligible weights) \citep{nguyen2014improving}. The ideas extend immediately to using the analogous SIR argument 
for obtaining additional posterior quantities of interest. 

\vspace{-5 truemm}

\section{Simulated data}\label{sim}

We conduct a simulation with $I=10,450$ individuals and $T=11$ capture occasions. We consider a constant capture probability and specify the survival probabilities to be a function of individual heterogeneity:
\begin{linenomath*}
\begin{eqnarray*}
p_{it+1} \: = \: p; & \mbox{and} & \text{logit }(\phi_{it}) \: = \:  \alpha+\epsilon_i, \nonumber
\end{eqnarray*}
\end{linenomath*}
for $i=1, \ldots, I$; $t = 1, \ldots, T-1$, where $\epsilon_i \sim N(0, \sigma^2)$. We set $p=0.13$, $\alpha=0.62$, and $\sigma=0.5$, corresponding to a realistic capture probability for many species, a median survival probability of 0.65 with lower and upper 2.5\% quantiles (0.41, 0.83). This is the same length of study as for the case study but for a reduced number of individuals and simpler model, so that we are able to analyse the full dataset using a standard Bayesian data augmentation approach for comparison. 

We used a stratified sampling approach, with strata defined to be the set of individuals released at time $t=1,\dots,T-1$ and observed for the final time at occasion $\tau = t,t+1,\dots,T$ (a total of 54  
strata). The number of individuals sampled from each strata was set equal to its observed proportion (rounded up to an integer). Within each strata, we uniformly selected the individual histories without replacement. To determine subsample size, we implemented a pilot-tuning stage using subsamples sizes between 5\%-30\%. Sample sizes $\ge 20\%$ had consistently similar subposterior distributions; whereas the subposterior distribution of subsamples $\le 10\%$, displayed much greater variability and level of particle depletion within the importance sampling step. Thus, we used a subsample size of 20\% (2,090 individuals) as a compromise between consistently similar subposterior distributions and reasonable computational cost. We simulated $M=100$ subsampled datasets.  
Finally, we specified the prior distributions: $p \sim$ {U}$(0,1)$, $\alpha \sim$ {N}$(0,10)$, and $\sigma \sim$ {U}$(0,10)$. 

For each subsampled dataset, we fitted the model using NIMBLE, specifying three independent MCMC chains, running each for 15,000 iterations, following a burn-in of 5,000 iterations. The simulations took approximately 10 minutes on an Intel$^{{\textregistered}}$Xeon$^{{\textregistered}}$
CPU E5-2683 v4 at 2.10 GHz and 64-bit Scientific Linux Mint 18.2 Sonya. For each subposterior, we thinned the sampled parameter values by 15 (i.e. retaining 1000 sampled values) and calculated their associated IS weights using a stratified MC approach with $N=100$ particles (this took approximately 4 minutes). Across the subsamples, the mean number of particles with non-negligible weight ($>0.001$) was 203, and ranged from 78-260. We used an SIR approach to obtain the associated 95\% symmetric credible intervals (CIs). For comparison, we also fitted the model to the full database directly using a Bayesian data augmentation approach. Due to the increased level of auto-correlation of the parameters (and posterior correlation between the random effect terms and associated variance), the simulations were run for 1 million iterations, with the first 100,000 discarded as burn-in (approximately 3 days to run). Table \ref{times} provides a summary comparison of the computational times for the different approaches. 

\vspace{0.5cm}

\begin{table}[H]
\begin{center}
\resizebox{13.5cm}{!}{
\begin{tabular}{l|c|c|c}
Computational time & MCMC iterations & Importance sampling weights & Total \\ 
\hline
\hline
Simulated data: subsampling approach & 10 minutes  & 4  minutes &  14 minutes \\
Simulated data: full data approach & 3 days & $-$ & 3 days\\
Case study: subsampling approach & 45 minutes & 29 minutes & 74 minutes\\
\hline
\end{tabular}
}
\caption{\label{times} Computational times (to nearest minute) for fitting the individual heterogeneity model to the simulated data and case study. For the simulated database 60,000 MCMC iterations are used with $N=100$ MC particles for 1000 sampled parameter values; and using the standard Bayesian data augmentation approach on the full dataset using 1 million MCMC iterations (to ensure convergence). For the case study 35,000 MCMC iterations are run and for the importance sampling step, a two-step approach is applied, using $N=25$ MC particles in Step 1 for 5000 sampled parameter values and $N=250$ particles in Step 2 retaining the top 500 ranked particles.} 
\end{center}
\end{table}

\begin{figure}[H]
\includegraphics[width=1\textwidth]{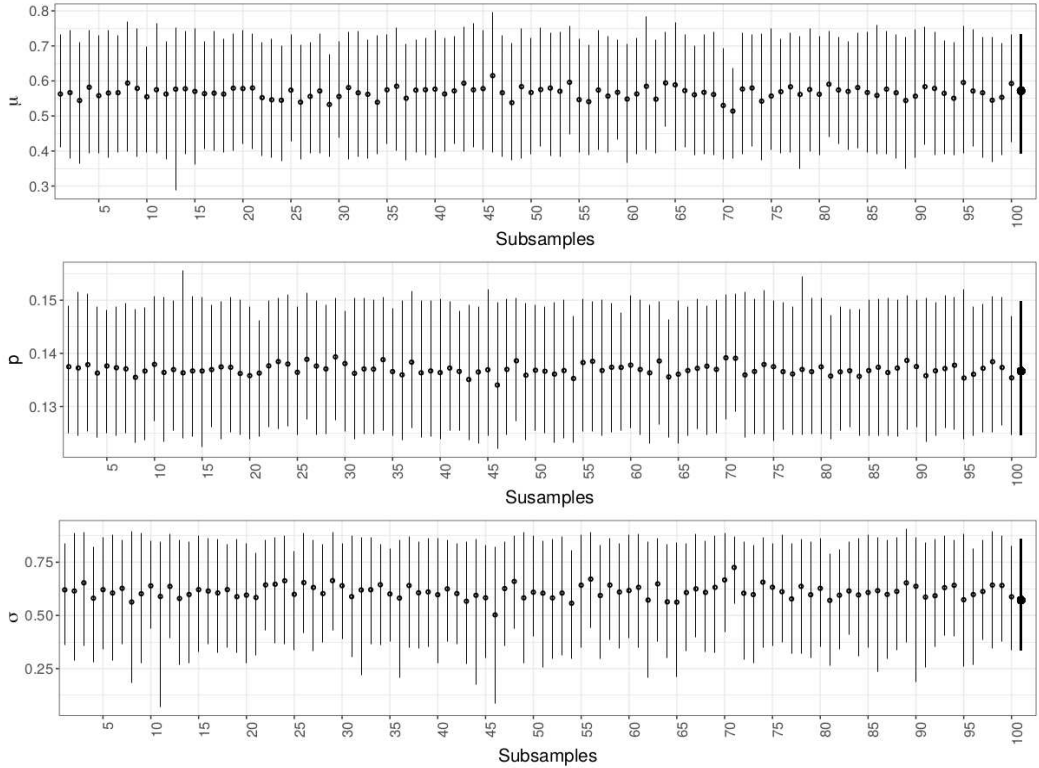}
\caption{\label{SIR_sub}Simulation study: Corrected posterior means and 95\% symmetric CIs for model parameters ($\mu$, $p$, and $\sigma$) and for the full simulated database (thick solid error bar).}

\end{figure}

The subposterior distributions where over-dispersed compared to the full posterior distribution, as expected. This can be seen in Table \ref{post_95CI} where we provide summary statistics of the lower and upper 2.5\% quantiles for the subposterior compared to (corrected) full posteriors across subsamples. The corresponding results for each (corrected) posterior for each subsampled dataset and associated estimate obtained from directly fitting the model to the full data are provided in Figure \ref{SIR_sub}. These posterior estimates are generally very similar to those obtained using the full data. (The corresponding subposteriors are provided in Web Appendix A in the Supplementary Material for each subsample). We combine the corrected posteriors across each subsample into a single estimate of the posterior. For simplicity, we assume an equal weight over subsamples, although using alternative weights gave essentially identical estimates. Table \ref{resamp} provides a summary of the associated posterior means, standard deviations and 95\% CIs. The combined estimate of the model parameters are very similar to those obtained from directly fitting the model to the full data; for all quantities displayed in Table \ref{resamp}, the estimates all differ by less than 1\%. However, the estimates obtained by our proposed approach are at a substantially reduced computationally cost. 
\vspace{0.5cm}


\begin{table}[H]
\centering
\resizebox{12cm}{!}{
\begin{tabular}{l|cc|cc}
&\multicolumn{2}{c}{Subposterior distribution} & \multicolumn{2}{c}{Full posterior distribution} \\
\hline
& Mean lower  & Mean upper & Mean lower & Mean upper \\
&2.5\% quantile & 2.5\% quantile & 2.5\% quantile & 2.5\% quantile \\
\hline
$\mu$ & 0.1740 & 0.8643 & 0.3935 & 0.7318 \\
$p$ & 0.1134 & 0.1658 & 0.1248 & 0.1499 \\
$\sigma$ & 0.2159 & 1.1677 & 0.3272 & 0.8565 \\
 \hline
\end{tabular}
}
\caption{\label{post_95CI} Simulation study: Mean lower and upper 2.5\% quantiles for the model parameters across the 100 subsamples for the subposterior distribution and full posterior distribution.}
\end{table}
\vspace{-3 truemm}
\begin{table}[H]
\centering
\begin{tabular}{l|ccc|ccc}
  & \multicolumn{3}{c}{Combined approach} &\multicolumn{3}{c}{Full simulated dataset} \\
  \hline
  & Mean & Sd & 95\% CI & Mean & Sd & 95\% CI \\ 
  \hline
$\mu$ & 0.5670 & 0.0887& [0.3915, 0.7361]& 0.5679 & 0.0879 & [0.3934, 0.7384] \\ 
 $p$ &0.1369& 0.0065&[0.1245, 0.1500]&0.1369 & 0.0065 & [0.1246, 0.1499] \\ 
  $\sigma$ & 0.6128 & 0.1379&[0.3179, 0.8613]&0.6118 & 0.1380 & [0.3149, 0.8621]\\ 
    \hline
\end{tabular}
\vspace{0.5cm}
\caption{\label{resamp} Simulation study: Posterior mean, standard deviation and 95\% symmetric CIs for the model parameters using the proposed importance sampling approach and combined across the 100 subsampled datasets and using a Bayesian data augmentation approach for the full simulated dataset.}
\end{table}

Following this ``proof-of-concept'' simulated dataset we apply the approach to the more challenging case study, where the model is more complex in terms of age and temporal dependencies in addition to the individual heterogeneity on the survival component. 

\vspace{-5 truemm}

\section{Case study: guillemots}\label{data}

We consider the case study described in Section \ref{case}. Given the number of ringed birds (28,930), the inclusion of individual heterogeneity on the survival probabilities is computationally challenging using the standard Bayesian data augmentation approach, even for relatively simple parameter dependence models. Incorporating additional biologically sensible parameter dependencies leads to added computational challenges. Motivated by \cite{Sarzo21}, and incorporating the known life cycle of guillemots, we consider an age-dependent model, where the survival and recapture probabilities have age structures: 1, 2, 3 and 4+, and 2, 3, 4, 5+, respectively. The survival probabilities are assumed to have additional temporal effects to reflect (unobserved) environmental heterogeneity over time, such as food availability, environmental conditions, etc. Mathematically, we let $a(i,t)$ denote the age of individual $i=1,\dots,I$ at time $t=1,\dots,T-1$, such that the parameters are of the form:
\begin{linenomath*}
\begin{eqnarray*}
p_{it+1} & = & p_{a(i,t+1)}; \mbox{ and } \mbox{logit } \phi_{it} = \alpha_{a(i,t)} + \beta_t + \epsilon_i, \mbox{ where } \epsilon_i \sim N(0,\sigma^2),
\end{eqnarray*}
\end{linenomath*}
for $t=1,\dots,T-1$ and $i=1, \ldots, I$. 
We specify vague prior distributions. For the temporal survival effects, we use a hierarchical distribution, such that $\beta_t \sim $ {N}$(\mu, \sigma)$, where $\mu \sim$ {N}$(0,10)$ and $\sigma \sim$ U$(0,10)$. For the age effect survival terms we set $\alpha_1=0$ (for identifiability) and $\alpha_a \sim $ {N}$(0,4)$, for $a=2, \ldots, 4+$. For the resighting probabilities, we specify $p_a \sim $ U$(0,1)$, for $a = 2, \ldots,5+$. Finally for the individual effects variance term, we set $\sigma_{\epsilon} \sim$ U$(0,2)$.

We apply the same subsampling scheme as in Section \ref{sim}, stratifying the histories based on initial and final capture times (54 possible strata). We subsampled $M=100$ datasets of sample size corresponding to 20\% of the database (i.e. 5,789 individuals). For each dataset, the model was fitted via NIMBLE, using 35,000 MCMC iterations, following a burn-in of 5,000 iterations (sampling random chains suggested that this was sufficient for convergence). Each MCMC simulation took approximately 45 minutes on an Intel$^{{\textregistered}}$Xeon$^{{\textregistered}}$
CPU E5-2683 v4 at 2.10 GHz and 64-bit Scientific Linux Mint 18.2 Sonya. Due to the increased complexity of this model we implemented a two-step stratified Monte Carlo approach for the importance sampling step. For Step 1, we thinned the MCMC sampled values by 70, providing 5000 sampled values, and for each of these calculated their (coarse) weights using $N=25$ MC particles. We retained the top 10\% (i.e. 500) sampled values and recalculated their weights using $N=250$ MC particles. This two-step MC approach took approximately 29 minutes per subsample.  
Table \ref{times} provides a summary of the computational times. To assess for convergence we repeated the two-step approach multiple times for a number of the subsamples. We consistently retained all the particles with non-negligible weight following Step 1, and obtained consistent weights for Step 2. 
The mean number of particles with a minimum weight of 0.0001 was 42 (range 5-97). The increased level of particle depletion (compared to the simulated data) is unsurprising given the increased dimension of the parameter space (18 parameters). 

\begin{figure}[H]
\includegraphics[width=1\textwidth]{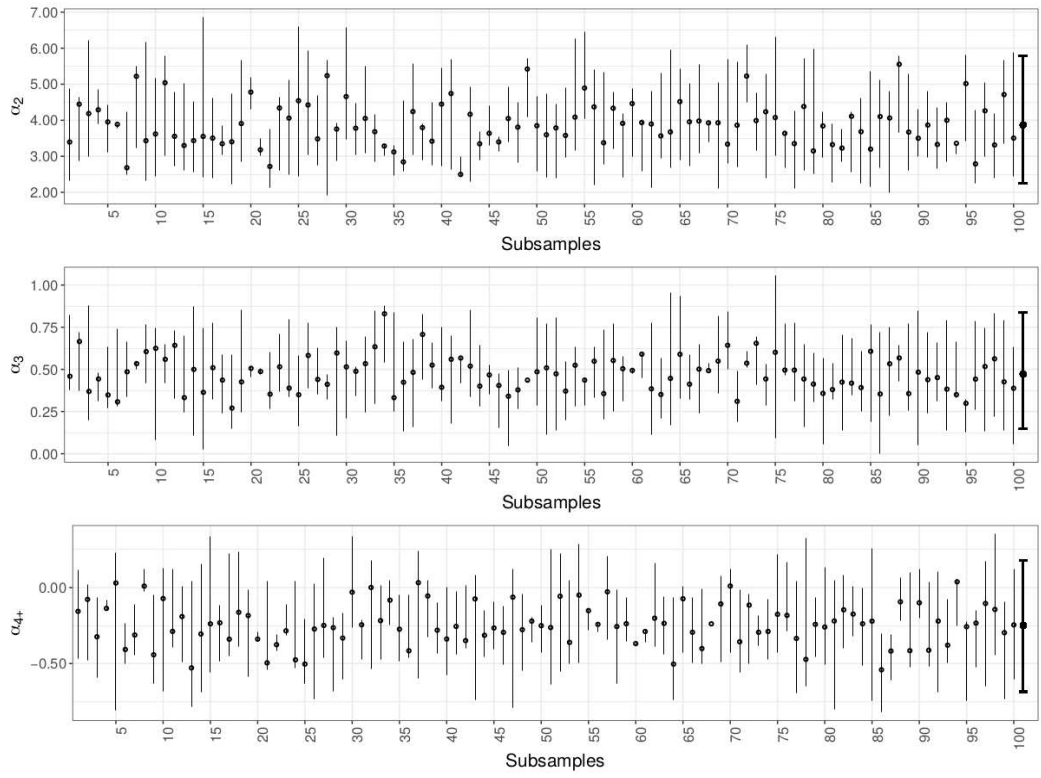}
\caption{\label{alphas}Case study: Corrected posterior means and 95\% symmetric CIs for $\alpha_a$ parameters for each subsample and combined across all subsamples (thick solid error bar) by age $a=2, \ldots, 4+$.}
\end{figure}

\begin{figure}[H]
\includegraphics[width=1\textwidth]{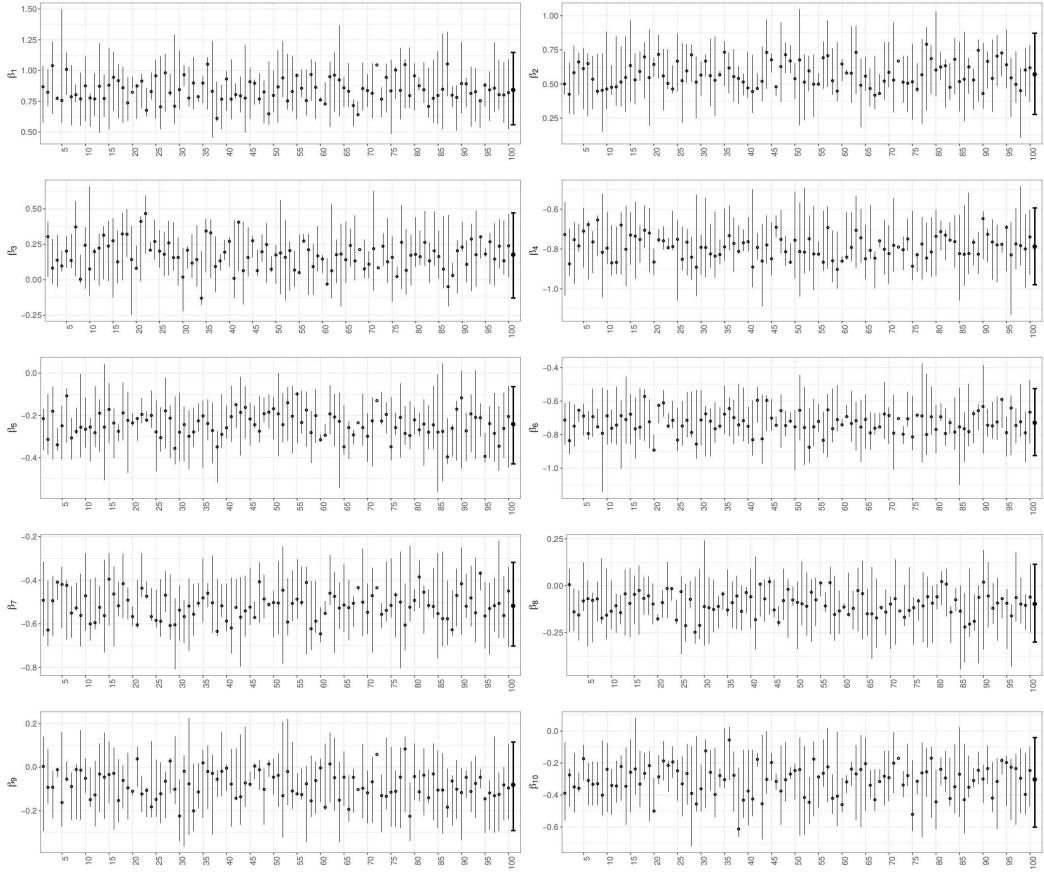}
\caption{\label{betas} Case study: Corrected posterior means and 95\% symmetric CIs for $\beta_t$ parameters for each subsample and combined across all subsamples (thick solid error bar) for occasions $t=1, \ldots, T-1$.}
\end{figure}

\begin{figure}[H]
\includegraphics[width=1\textwidth]{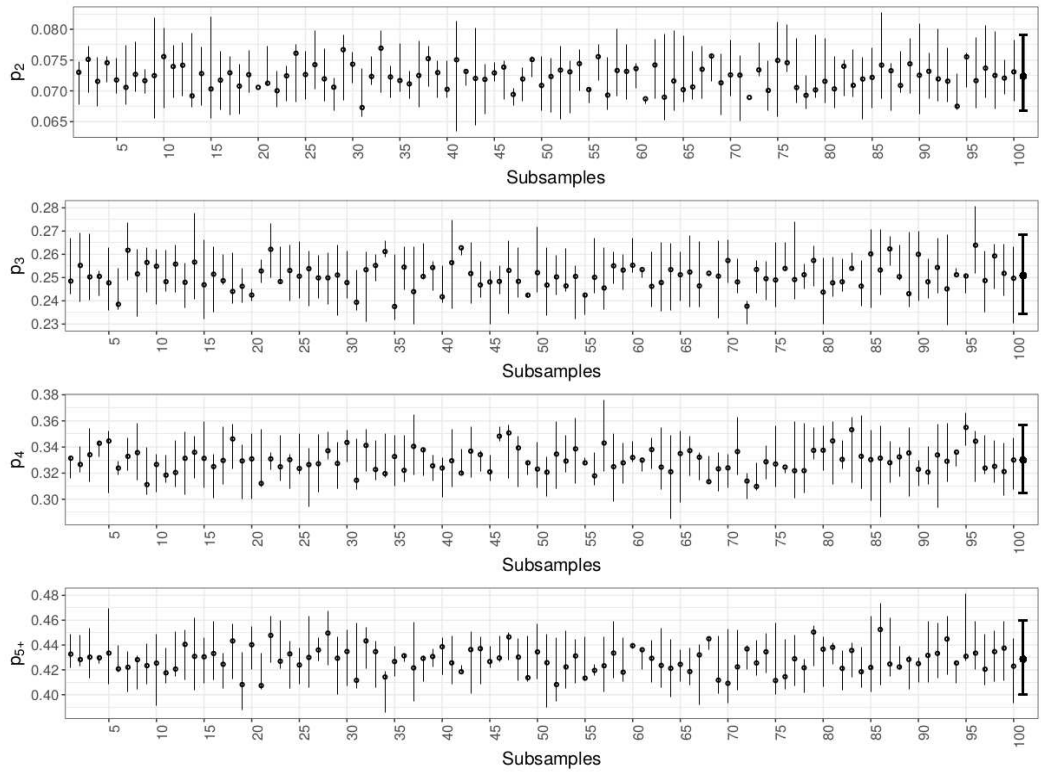}
\caption{\label{recap}Case study: Corrected posterior means and 95\% symmetric CIs for recapture probabilities for each subsample and combined across all subsamples (thick solid error bar) by age ($a=2, \ldots, 5+$).}
\end{figure}

\begin{figure}[H]
\begin{center}
\includegraphics[width=0.8\textwidth]{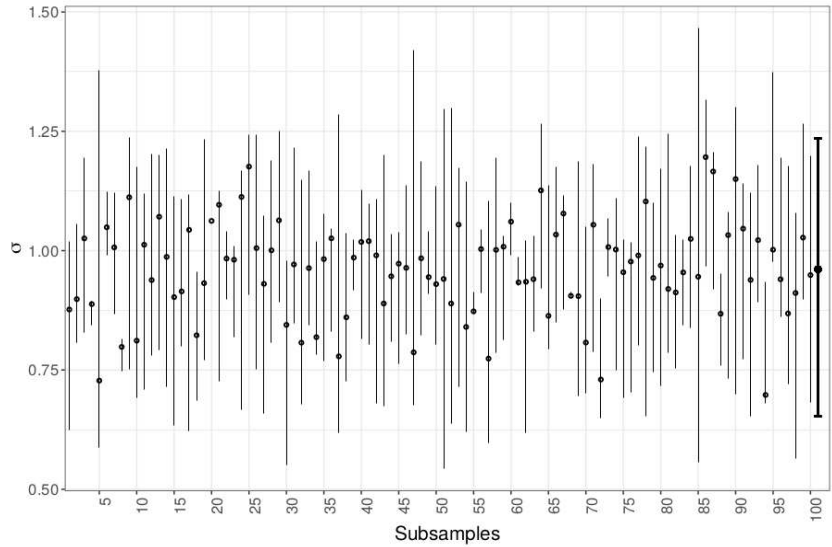}
\caption{\label{sigma}Case study: Corrected posterior means and 95\% symmetric CIs for the variance of the individual effects ($\sigma_{\epsilon}$) for each subsample and combined across all subsamples (thick solid error bar).}
\end{center}
\end{figure}

Table \ref{RMSE} provides the (corrected) posterior mean and standard deviation (SD) for each  parameter combined over the subsamples; while Figures \ref{alphas} to \ref{sigma} provide the estimated (corrected) posterior mean and 95\% CIs for each subsample, and combined across all subsamples. There is some variability of the posterior distribution per subsample (though generally overlapping), which is unexpected given the reduced effective sample sizes. However, we are able to obtain an estimate of the posterior by combining the subsample estimates, immediately increasing the sample size and providing increased accuracy. To investigate the robustness of this approach, we randomly selected 25 and 50 samples (without replacement) of the estimated posterior distributions obtained from the full set of subsamples and calculated the associated posterior mean and SD. We repeated this a total of 100 times and calculated the corresponding root mean square error of the given posterior summary statistics, compared to the estimate obtained using all subsamples. The results are given in Table \ref{RMSE}, which suggests that the estimates of the posterior summary statistics are fairly robust when combining across subsamples, even when some individual subsamples lead to low effective sample sizes. As expected there is smaller variability when using 50 subsamples compared to 25. 

\vspace{0.2cm}
\begin{table}[H]
\centering
\begin{tabular}{l|cc|cc|cc}
&\multicolumn{2}{c}{$100$ subsamples}&\multicolumn{2}{c}{$50$ subsamples} & \multicolumn{2}{c}{$25$ subsamples} \\
  \hline
 & \multicolumn{2}{c}{Posterior} & \multicolumn{2}{c}{RMSE} & \multicolumn{2}{c}{RMSE}  \\ 
 & Mean & SD & Mean & SD & Mean & SD \\
  \hline
$\beta_1$ & 0.842 & 0.156& 0.010 & 0.008 & 0.017 & 0.012 \\ 
$\beta_2$ & 0.571& 0.161& 0.009 & 0.008 & 0.016 & 0.011 \\ 
$\beta_3$ & 0.177& 0.158& 0.013 & 0.010 & 0.019 & 0.015  \\ 
$\beta_4$ & -0.788& 0.103& 0.005 & 0.006 & 0.010 & 0.009  \\ 
$\beta_5$ & -0.242& 0.101& 0.009 & 0.005 & 0.011 & 0.007 \\ 
$\beta_6$ & -0.729& 0.105& 0.006 & 0.005 & 0.011 & 0.008 \\ 
$\beta_7$ & -0.518& 0.106& 0.006 & 0.007 & 0.009 & 0.009 \\ 
$\beta_8$ & -0.097 & 0.107& 0.006 & 0.006 & 0.011 & 0.008 \\ 
$\beta_9$ & -0.081& 0.104& 0.009 & 0.005 & 0.012 & 0.009 \\ 
$\beta_{10}$ & -0.303 & 0.147& 0.012 & 0.007 & 0.019 & 0.012 \\
   \hline
$\alpha_2$ & 3.871 & 0.930& 0.089 & 0.133 & 0.143 & 0.149  \\ 
$\alpha_3$ & 0.472 & 0.176& 0.011 & 0.010 & 0.020 & 0.013  \\ 
$\alpha_{4+}$ & -0.248 & 0.223& 0.018 & 0.015 & 0.025 & 0.026  \\
   \hline
$p_2$ & 0.072 & 0.003&  0.015 & 0.012 & 0.019 & 0.015  \\ 
$p_3$ & 0.251 & 0.009& 0.023 & 0.019 & 0.028 & 0.025  \\ 
$p_4$ & 0.330 & 0.014& 0.033 & 0.024 & 0.037 & 0.030  \\ 
$p_{5+}$ & 0.429 & 0.015& 0.029 & 0.025 & 0.038 & 0.029  \\  
  \hline
  $\sigma$ & 0.957 & 0.130& 0.017 & 0.021 & 0.018 & 0.022 \\
  \hline
\end{tabular}
\vspace{0.4cm}
\caption{\label{RMSE}Case study: Posterior mean and standard deviation (SD) of the model parameters for the combined full posterior distribution (using 100 subsampled datasets from the full dataset) and associated root mean square error (RMSE) for 50 and 25 randomly sampled posterior distribution (without replacement). }
\end{table}

From Figure \ref{sigma}, there appears to be a substantial random effect variance component (on the logit scale), with the posterior mean of $\sigma$ equal to $0.96$, with 95\% CI $[0.65, 1.24]$. This suggests a reasonable amount of unobserved individual heterogeneity present, unexplained by the individual age effects. This may, for example, be representative of inherent differences in individual quality or condition. We compare the posterior estimates of the parameters with the model omitting the individual heterogeneity component in Web Appendix B in the Supplementary Material. We note that the inclusion of the individual heterogeneity leads to similar parameter estimates for the survival temporal effects and capture probabilities, but with substantially larger credible intervals for the survival probabilities across ages. 

\vspace{-2 truemm}

\section{Discussion}\label{disc}

Advances in computational resources and readily available computer packages have permitted the fitting of more complex models to real data across the breadth of the scientific community. However, computational limitations remain for many real applications, particularly as increasing amounts of data become available. In such circumstances, applying standard computational algorithms may become prohibitive. In this paper, we consider fitting (continuous) individual heterogeneity models to a large capture-recapture dataset, for which using the standard Bayesian data augmentation approach is impractical. 

We propose exploring the posterior distribution of a subsample of the data that is then corrected via importance sampling such that we obtain an estimate of the posterior distribution of the full dataset. The approach is embarrassingly parallelisable in two aspects: in terms of the multiple subsamples, and calculating the importance sampling (unnormalised) weights of each subsample. Further, the algorithm can be easily implemented requiring essentially only one bespoke function corresponding to a Monte Carlo estimate of the probability of a given capture history, and black-box MCMC samplers (such as in JAGS/NIMBLE). 

For the guillemot case study, we consider an individual heterogeneity effect on the survival probability, for which using the standard Bayesian data augmentation approach becomes infeasibly slow. However, using our proposed approach, we were able to obtain an estimate of the posterior distribution of the full dataset using NIMBLE in the order of magnitude of an hour, considering 20\% of the capture histories within the subsampled datasets. Moreover, multiple subsamples can be run simultaneously, with the limiting factor simply the number of computing cores available, and combined to obtain more robust and reliable results. The corresponding results estimated the posterior mean of the random effect standard deviation to be equal to 0.96 (where the random effect is on the logistic scale), suggesting a reasonably high level of heterogeneity present in the (\textit{apparent}) survival probabilities of individuals.

The proposed algorithm is more generally applicable to intractable likelihood problems of large datasets. There are a number of practical implementation issues to be considered for such problems, including, for example, the ``optimal'' sub-sampling size and/or subsampling strata to be used (in order to minimize the mismatch between subposterior and the  posterior distribution of the full dataset). The efficiency of the approach relies on the subposterior being similar to the full posterior, to minimise particle depletion and reduce the effective sample size. Thus an additional step that may be considered is the inclusion of an accept/reject step following the simulation of a subsampled dataset, retaining the subsample only if it has similar enough ``properties'' to the full data (with the aim that this increases the probability that the posteriors are similar). For example, such properties could be a function of (scaled) sufficient statistics of the given dataset. Alternatively to decrease the particle depletion, the selection of sampled MCMC parameter values to be used may be considered further, considering the autocorrelation of the parameter values and/or using a multi-step algorithm for selecting the set of parameter values, following the calculation of the weights of a given set of parameter values in an initial step. Finally, other potential extensions that may be explored within the MC step include the specification of the threshold to determine the samples to retain for the second step for computational efficiency whilst retaining high precision where the threshold may depend on the variability of the (coarse) weights, or the use of a nonlinear transformation of the importance weights in order to reduce particle depletion (e.g., as in \citealp{ionides2008truncated,vehtari2015pareto}). These areas are the focus of current research. 

\vspace{-3 truemm}

\section*{Acknowledgements}

We thank the Baltic Seabird Project for making the data available and the large number of field workers and volunteers at Stora Karls\"o. Field work on Stora Karls\"o has been made possible through a long-term engagement in the Baltic Seabird project by WWF Sweden. 
RK was supported by the Leverhulme research fellowship RF-2019-299. BS was supported by Margarita Salas fellowship from Ministry of Universities-University of Valencia (MS21-013). For the purpose of open access, the author has applied a Creative Commons Attribution (CC BY) licence to any Author Accepted Manuscript version arising from this submission.

\vspace{-4 truemm}

\section*{Supporting Information}

Web Appendices A and B referenced in Section \ref{case} are available at the Supplementary Material. The code used in Section \ref{sim} for the simulation study is available at: \url{https://github.com/sarzoblanca/King-Sarzo-and-Elvira.-2022.-When-Worlds-Collide}. 

\vspace{0.5cm}

\bibliographystyle{chicago}
\doublespacing

\vspace{-5 truemm}

\bibliography{references.bib}

\begin{thebibliography}{}

\bibitem[\protect\citeauthoryear{Andrieu, Doucet, and Holenstein}{Andrieu
  et~al.}{2010}]{Andrieu_2010}
Andrieu, C., A.~Doucet, and R.~Holenstein (2010).
\newblock Particle {Markov} chain {Monte} {Carlo} methods.
\newblock {\em Journal of the Royal Statistical Society: Series B\/}~{\em 72},
  269--342.

\bibitem[\protect\citeauthoryear{Andrieu and Roberts}{Andrieu and
  Roberts}{2009}]{Andrieu_2009}
Andrieu, C. and G.~O. Roberts (2009).
\newblock The pseudo-marginal approach for efficient {Monte} {Carlo}
  {Computations}.
\newblock {\em The Annals of Statistics\/}~{\em 37}, 697--725.

\bibitem[\protect\citeauthoryear{Bardenet, Doucet, and Holmes}{Bardenet
  et~al.}{2017}]{bar17}
Bardenet, R., A.~Doucet, and C.~Holmes (2017).
\newblock On {M}arkov chain {M}onte {C}arlo methods for tall data.
\newblock {\em Journal of Machine Learning Research\/}~{\em 18\/}(1),
  1515–1557.

\bibitem[\protect\citeauthoryear{Brooks, Gelman, Jones, and Meng}{Brooks
  et~al.}{2011}]{mcmc}
Brooks, S.~P., A.~Gelman, G.~Jones, and X.~Meng (Eds.) (2011).
\newblock {\em Handbook of Markov Chain Monte Carlo; Methods and Applications}.
\newblock CRC Press.

\bibitem[\protect\citeauthoryear{Carpenter, Gelman, Hoffman, Lee, Goodrich,
  Bentancourt, Brubaker, Guo, and Riddell}{Carpenter et~al.}{2017}]{carpenter}
Carpenter, B., A.~Gelman, M.~D. Hoffman, D.~Lee, B.~Goodrich, M.~Bentancourt,
  M.~Brubaker, J.~Guo, and A.~Riddell (2017).
\newblock Stan: {A} probabilistic programming language.
\newblock {\em Journal of Statistical Software\/}~{\em 76\/}(1).

\bibitem[\protect\citeauthoryear{Coull and Agresti}{Coull and
  Agresti}{1999}]{coua99}
Coull, B.~A. and A.~Agresti (1999).
\newblock The use of mixed logit models to reflect heterogeneity in
  capture-recapture studies.
\newblock {\em Biometrics\/}~{\em 55}, 294--301.

\bibitem[\protect\citeauthoryear{{de Valpine}}{{de Valpine}}{2002}]{dev02}
{de Valpine}, P. (2002).
\newblock Review of methods for fitting time-series models with process and
  observation error and likelihood calculations for nonlinear, non-{G}aussian
  state-space models.
\newblock {\em Bulletin of Marine Science\/}~{\em 70}, 455--471.

\bibitem[\protect\citeauthoryear{{de Valpine}}{{de Valpine}}{2004}]{dev04}
{de Valpine}, P. (2004).
\newblock Monte {C}arlo state-space likelihoods by weighted posterior kernel
  density estimation.
\newblock {\em Journal of the American Statistical Association\/}~{\em 99},
  523--534.

\bibitem[\protect\citeauthoryear{{de Valpine}, Turek, Paciorek,
  Anderson-Bergman, Lang, and Bodik}{{de Valpine} et~al.}{2017}]{nimble}
{de Valpine}, P., D.~Turek, C.~Paciorek, C.~Anderson-Bergman, D.~Lang, and
  R.~Bodik (2017).
\newblock Programming with models: {W}riting statistical algorithms for general
  model structures with {NIMBLE}.
\newblock {\em Journal of Computational and Graphical Statistics\/}~{\em 26},
  403--413.

\bibitem[\protect\citeauthoryear{Douc, Guillin, Marin, and Robert}{Douc
  et~al.}{2007}]{Douc07}
Douc, R., A.~Guillin, J.~M. Marin, and C.~P. Robert (2007).
\newblock Minimum variance importance sampling via population {M}onte {C}arlo.
\newblock {\em ESAIM: Probability and Statistics\/}~{\em 11}, 427--447.

\bibitem[\protect\citeauthoryear{Elvira and Martino}{Elvira and
  Martino}{2021}]{elvm21}
Elvira, V. and L.~Martino (2021).
\newblock Advances in importance sampling.
\newblock {\em Wiley StatsRef: Statistics Reference Online\/}, 1--14.

\bibitem[\protect\citeauthoryear{Francis and Saurola}{Francis and
  Saurola}{2009}]{Francis09}
Francis, C.~M. and P.~Saurola (2009).
\newblock Estimating demographic parameters from complex data sets: {A}
  comparison of bayesian hierarchical and maximum-likelihood methods for
  estimating survival probabilities of tawny owls, \textit{Strix aluco} in
  {F}inland.
\newblock In D.~L. Thomson, E.~G. Cooch, and M.~J. Conroy (Eds.), {\em Modeling
  Demographic Processes In Marked Populations}, pp.\  617--637. Boston, MA:
  Springer.

\bibitem[\protect\citeauthoryear{Gelman, Carlin, Stern, and Rubin}{Gelman
  et~al.}{2014}]{gelman2014bayesian}
Gelman, A., J.~B. Carlin, H.~S. Stern, and D.~B. Rubin (2014).
\newblock {\em Bayesian {D}ata {A}nalysis}, Volume~2.
\newblock Chapman \& Hall/CRC Boca Raton, FL, USA.

\bibitem[\protect\citeauthoryear{Gimenez, Bonner, King, Parker, Brooks,
  Jamieson, Grosbois, Morgan, and Thomas}{Gimenez et~al.}{2009}]{gimenez09}
Gimenez, O., S.~Bonner, R.~King, R.~A. Parker, S.~P. Brooks, L.~E. Jamieson,
  V.~Grosbois, B.~J.~T. Morgan, and L.~Thomas (2009).
\newblock Win{BUGS} for population ecologists: {B}ayesian modelling using
  {M}arkov chain {M}onte {C}arlo ({MCMC}) methods.
\newblock In D.~L. Thomson, E.~G. Cooch, and M.~J. Conroy (Eds.), {\em Modeling
  Demographic Processes In Marked Populations}, pp.\  885--918. Boston, MA:
  Springer.

\bibitem[\protect\citeauthoryear{Gimenez, Cam, and J-M.}{Gimenez
  et~al.}{2017}]{gimcg17}
Gimenez, O., E.~Cam, and G.~J-M. (2017).
\newblock Individual heterogeneity and capture–recapture models: what, why
  and how?
\newblock {\em Oikos\/}~{\em 127}, 664--686.

\bibitem[\protect\citeauthoryear{Gimenez and Choquet}{Gimenez and
  Choquet}{2010}]{gimenez10}
Gimenez, O. and R.~Choquet (2010).
\newblock Individual heterogeneity in studies on marked animals using numerical
  integration: capture-recapture mixed models.
\newblock {\em Ecology\/}~{\em 91}, 951--957.

\bibitem[\protect\citeauthoryear{Gimenez, Rossi, Choquet, Dehais, Doris,
  Varella, Vila, and Pradel}{Gimenez et~al.}{2007}]{gimenez2007state}
Gimenez, O., V.~Rossi, R.~Choquet, C.~Dehais, B.~Doris, H.~Varella, J.~P. Vila,
  and R.~Pradel (2007).
\newblock State-space modelling of data on marked individuals.
\newblock {\em Ecological Modelling\/}~{\em 206}, 431--438.

\bibitem[\protect\citeauthoryear{Hankin, Mohr, and Newman}{Hankin
  et~al.}{2019}]{hanmn19}
Hankin, D., M.~Mohr, and K.~Newman (2019).
\newblock {\em Sampling Theory}.
\newblock Oxford University Press.

\bibitem[\protect\citeauthoryear{Herliansyah, King, and King}{Herliansyah
  et~al.}{2022}]{herkk21}
Herliansyah, R., R.~King, and S.~E. King (2022).
\newblock Laplace approximations for individual heterogeneity capture-recapture
  models.
\newblock {\em Journal of Agricultural, Biological, and Environmental
  Statistics\/}.
\newblock in press.

\bibitem[\protect\citeauthoryear{Hestbeck, Nichols, and Malecki}{Hestbeck
  et~al.}{1991}]{Hestbeck91}
Hestbeck, J.~B., J.~D. Nichols, and R.~A. Malecki (1991).
\newblock Estimates of movement and site fidelity using mark-resight data of
  wintering canada geese.
\newblock {\em Ecology\/}~{\em 72}, 523--533.

\bibitem[\protect\citeauthoryear{Ionides}{Ionides}{2008}]{ionides2008truncated}
Ionides, E.~L. (2008).
\newblock Truncated importance sampling.
\newblock {\em Journal of Computational and Graphical Statistics\/}~{\em
  17\/}(2), 295--311.

\bibitem[\protect\citeauthoryear{K{\'e}ry and Schaub}{K{\'e}ry and
  Schaub}{2011}]{kery2011bayesian}
K{\'e}ry, M. and M.~Schaub (2011).
\newblock {\em Bayesian Population Analysis using WinBUGS: A hierarchical
  perspective}.
\newblock Academic Press.

\bibitem[\protect\citeauthoryear{King}{King}{2014}]{king14}
King, R. (2014).
\newblock Statistical {E}cology.
\newblock {\em Annual Review of Statistics and its Application\/}~{\em 1},
  401--426.

\bibitem[\protect\citeauthoryear{King and Brooks}{King and
  Brooks}{2008}]{kinb08}
King, R. and S.~P. Brooks (2008).
\newblock On the {Bayesian} {estimation} of a {closed} {population} {size} in
  the {presence} of {heterogeneity} and {model} {uncertainty}.
\newblock {\em Biometrics\/}~{\em 64}, 816--824.

\bibitem[\protect\citeauthoryear{King, McClintock, Kidney, and Borchers}{King
  et~al.}{2016}]{kinmkb16}
King, R., B.~T. McClintock, D.~Kidney, and D.~Borchers (2016).
\newblock Capture-recapture abundance estimation using a semi-complete data
  likelihood approach.
\newblock {\em The Annals of Applied Statistics\/}~{\em 10}, 264--285.

\bibitem[\protect\citeauthoryear{King, Morgan, Giménez, and Brooks}{King
  et~al.}{2010}]{king2010bayesian}
King, R., B.~J.~T. Morgan, O.~Giménez, and S.~P. Brooks (2010).
\newblock {\em Bayesian {A}nalysis for {P}opulation {E}cology}.
\newblock CRC Press.

\bibitem[\protect\citeauthoryear{Luengo, Martino, Elvira, and Bugallo}{Luengo
  et~al.}{2018}]{luengo2018efficient}
Luengo, D., L.~Martino, V.~Elvira, and M.~Bugallo (2018).
\newblock Efficient linear fusion of partial estimators.
\newblock {\em Digital Signal Processing\/}~{\em 78}, 265--283.

\bibitem[\protect\citeauthoryear{Lunn, Thomas, Best, and Spiegelhalter}{Lunn
  et~al.}{2000}]{lunn2000}
Lunn, D.~J., A.~Thomas, N.~Best, and D.~Spiegelhalter (2000).
\newblock Win{BUGS}: a {{Bayesian}} modelling framework: concepts, structure,
  and extensibility.
\newblock {\em Statistics and Computing\/}~{\em 10}, 325--337.

\bibitem[\protect\citeauthoryear{McCrea and Morgan}{McCrea and
  Morgan}{2015}]{McCrea}
McCrea, R. and B.~J.~T. Morgan (2015).
\newblock {\em Analysis of Capture-Recapture Data}.
\newblock CRC Press.

\bibitem[\protect\citeauthoryear{Nguyen, Septier, Peters, and Delignon}{Nguyen
  et~al.}{2014}]{nguyen2014improving}
Nguyen, T. L.~T., F.~Septier, G.~W. Peters, and Y.~Delignon (2014).
\newblock Improving {SMC} sampler estimate by recycling all past simulated
  particles.
\newblock In {\em Statistical Signal Processing (SSP), 2014 IEEE Workshop on},
  pp.\  117--120. IEEE.

\bibitem[\protect\citeauthoryear{Olsson and Hentati-Sundberg}{Olsson and
  Hentati-Sundberg}{2017}]{olsson2017}
Olsson, O. and J.~Hentati-Sundberg (2017).
\newblock Population trends and status of four seabird species (\textit{Uria
  aalge}, \textit{Alca torda}, \textit{Larus fuscus}, \textit{Larus
  argentatus}) at {S}tora {K}arls\"{o} in the {B}altic sea.
\newblock {\em Ornys Svecica\/}~{\em 27}, 64--93.

\bibitem[\protect\citeauthoryear{Owen}{Owen}{2013}]{mcbook}
Owen, A.~B. (2013).
\newblock {\em Monte Carlo theory, methods and examples}.

\bibitem[\protect\citeauthoryear{Pledger}{Pledger}{2000}]{ple00}
Pledger, S. (2000).
\newblock Unified maximum likelihood estimates for closed capture-recapture
  models using mixtures.
\newblock {\em Biometrics\/}~{\em 56}, 434--442.

\bibitem[\protect\citeauthoryear{Pledger, Pollock, and Norris}{Pledger
  et~al.}{2003}]{plepn03}
Pledger, S., K.~H. Pollock, and J.~L. Norris (2003).
\newblock Open capture-recapture models with heterogeneity. {I}
  {C}ormack-{J}olly-{S}eber.
\newblock {\em Biometrics\/}~{\em 59}, 786--794.

\bibitem[\protect\citeauthoryear{Plummer}{Plummer}{2003}]{plummer2003jags}
Plummer, M. (2003).
\newblock {JAGS}: A program for analysis of {B}ayesian graphical models using
  {G}ibbs sampling.
\newblock In {\em Proceedings of the 3rd international workshop on distributed
  statistical computing}, Volume 124, pp.\  1--9.

\bibitem[\protect\citeauthoryear{Robert, Elvira, Tawn, and Wu}{Robert
  et~al.}{2018}]{robert2018accelerating}
Robert, C.~P., V.~Elvira, N.~Tawn, and C.~Wu (2018).
\newblock Accelerating {MCMC} algorithms.
\newblock {\em Wiley Interdisciplinary Reviews: Computational
  Statistics\/}~{\em 10}, 1--14.

\bibitem[\protect\citeauthoryear{Royle}{Royle}{2008}]{royle2008modeling}
Royle, J.~A. (2008).
\newblock Modeling individual effects in the {C}ormack--{J}olly--{S}eber model:
  {A} state--space formulation.
\newblock {\em Biometrics\/}~{\em 64}, 364--370.

\bibitem[\protect\citeauthoryear{Sarzo, Armero, Conesa, Hentati-Sundberg, and
  Olsson}{Sarzo et~al.}{2019}]{Sarzo19}
Sarzo, B., C.~Armero, D.~Conesa, J.~Hentati-Sundberg, and O.~Olsson (2019).
\newblock Bayesian immature survival analysis of the largest colony of {C}ommon
  murre \textit{{U}ria aalge} in the {B}altic sea.
\newblock {\em Waterbirds\/}~{\em 42}, 304--313.

\bibitem[\protect\citeauthoryear{Sarzo, King, Conesa, and
  Hentati-Sundberg}{Sarzo et~al.}{2021}]{Sarzo21}
Sarzo, B., R.~King, D.~Conesa, and J.~Hentati-Sundberg (2021).
\newblock Correcting bias in survival probabilities for partially monitored
  populations via integrated models.
\newblock {\em Journal of Agricultural, Biological and Environmental
  Statistics\/}~{\em 26\/}(2), 200--219.

\bibitem[\protect\citeauthoryear{Seber and Schofield}{Seber and
  Schofield}{2019}]{seber19}
Seber, G. and M.~Schofield (2019).
\newblock {\em Capture-Recapture: Parameter Estimation for Open Animal
  Populations}.
\newblock Springer.

\bibitem[\protect\citeauthoryear{Tokdar and Kass}{Tokdar and
  Kass}{2010}]{tokk10}
Tokdar, S.~T. and R.~E. Kass (2010).
\newblock Importance sampling: a review.
\newblock {\em Wiley Interdisciplinary Reviews: Computational
  Statistics\/}~{\em 2}, 54--60.

\bibitem[\protect\citeauthoryear{Tran, Scharth, Pitt, and Kohn}{Tran
  et~al.}{2016}]{tran16}
Tran, M.-N., M.~Scharth, M.~K. Pitt, and R.~Kohn (2016).
\newblock Importance sampling squared for {B}ayesian inference in latent
  variable models.
\newblock {\em arXiv:1309.3339\/}.

\bibitem[\protect\citeauthoryear{van~de Schoot, Depaoli, King, Kramer,
  M{\"a}rtens, Tadesse, Vannucci, Gelman, Veen, Willemsen, and Yau}{van~de
  Schoot et~al.}{2021}]{van21}
van~de Schoot, R., S.~Depaoli, R.~King, B.~Kramer, K.~M{\"a}rtens, M.~G.
  Tadesse, M.~Vannucci, A.~Gelman, D.~Veen, J.~Willemsen, and C.~Yau (2021).
\newblock {B}ayesian statistics and modelling.
\newblock {\em Nature Reviews Methods Primers\/}~{\em 1}, 1--26.

\bibitem[\protect\citeauthoryear{Vehtari, Simpson, Gelman, Yao, and
  Gabry}{Vehtari et~al.}{2015}]{vehtari2015pareto}
Vehtari, A., D.~Simpson, A.~Gelman, Y.~Yao, and J.~Gabry (2015).
\newblock Pareto smoothed importance sampling.
\newblock {\em arXiv:1507.02646\/}.

\end{thebibliography}


\begin{thebibliography}{}

\bibitem[\protect\citeauthoryear{Sarzo, King, Conesa, and
  Hentati-Sundberg}{Sarzo et~al.}{2021}]{Sarzo21}
Sarzo, B., R.~King, D.~Conesa, and J.~Hentati-Sundberg (2021).
\newblock Correcting bias in survival probabilities for partially monitored
  populations via integrated models.
\newblock {\em Journal of Agricultural, Biological and Environmental
  Statistics\/}~{\em 26\/}(2), 200--219.

\end{thebibliography}

\end{document}



\begin{center}

\textbf{\LARGE{Supplementary Material for: \\
Large Data and (Not Even Very) Complex Ecological Models: When Worlds Collide}  }

~

\large{Ruth King$^1$, Blanca Sarzo$^{1,2}$, and V\'ictor Elvira$^1$ \\
    $^1$School of Mathematics and Maxwell Institute for Mathematical Sciences, University of Edinburgh, \\
    Edinburgh, UK \\
    $^2$Department of Microbiology and Ecology, University of Valencia, Valencia, Spain \\
}
\end{center}

\section*{Web Appendix A}

\begin{figure}[H]
\includegraphics[width=1\textwidth]{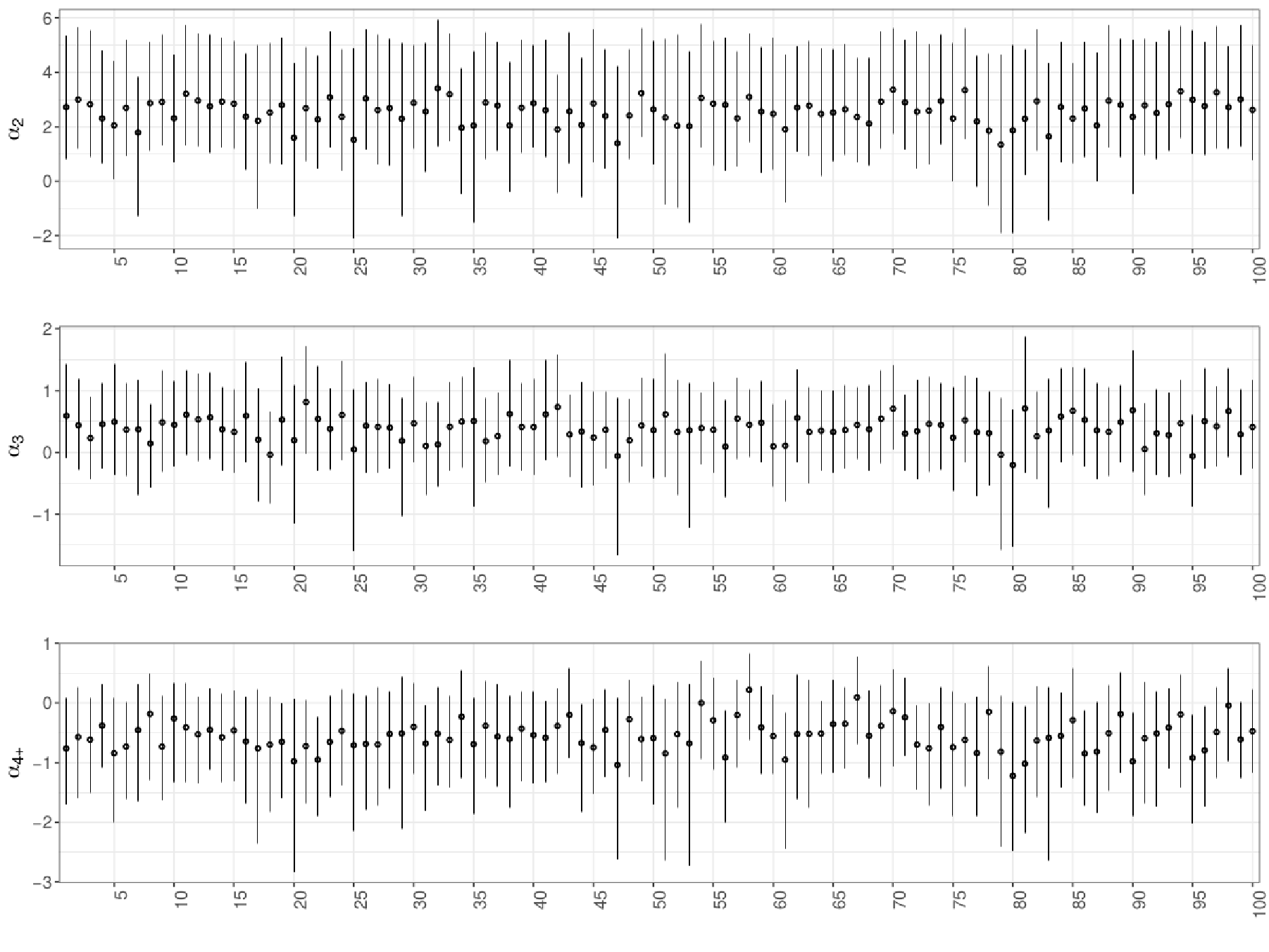}
\caption{\label{alpha}Case study: Subposterior means and 95\% symmetric CIs for $\alpha_a$ parameters for each subsample for ages $a=2, \ldots, 4+$.}
\end{figure}

\begin{figure}[H]
\includegraphics[width=1\textwidth]{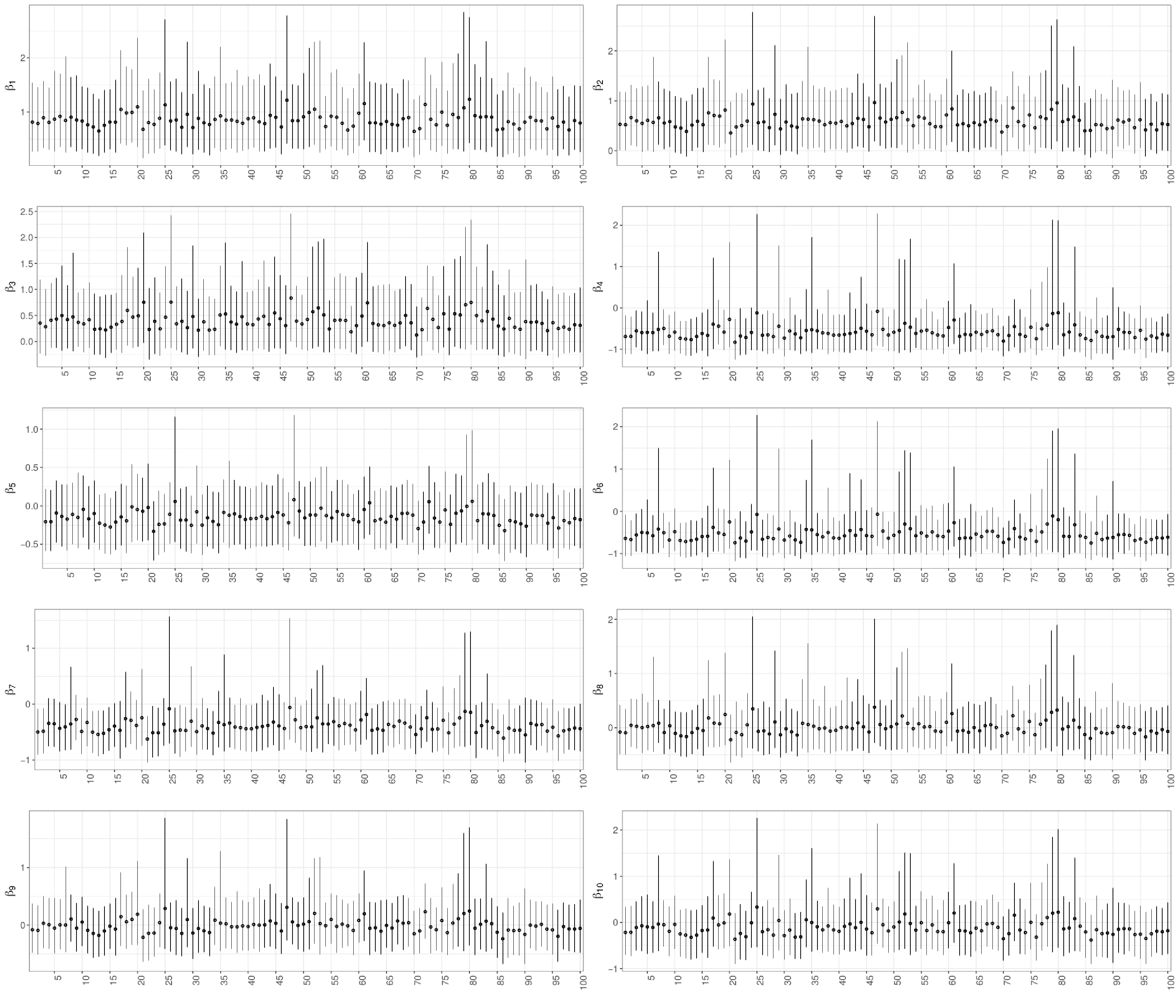}
\caption{\label{betas}Case study: Subposterior means and 95\% symmetric CIs for $\beta_t$ parameters for each subsample for occasions $t=1, \ldots, T-1$.}
\end{figure}

\begin{figure}[H]
\includegraphics[width=1\textwidth]{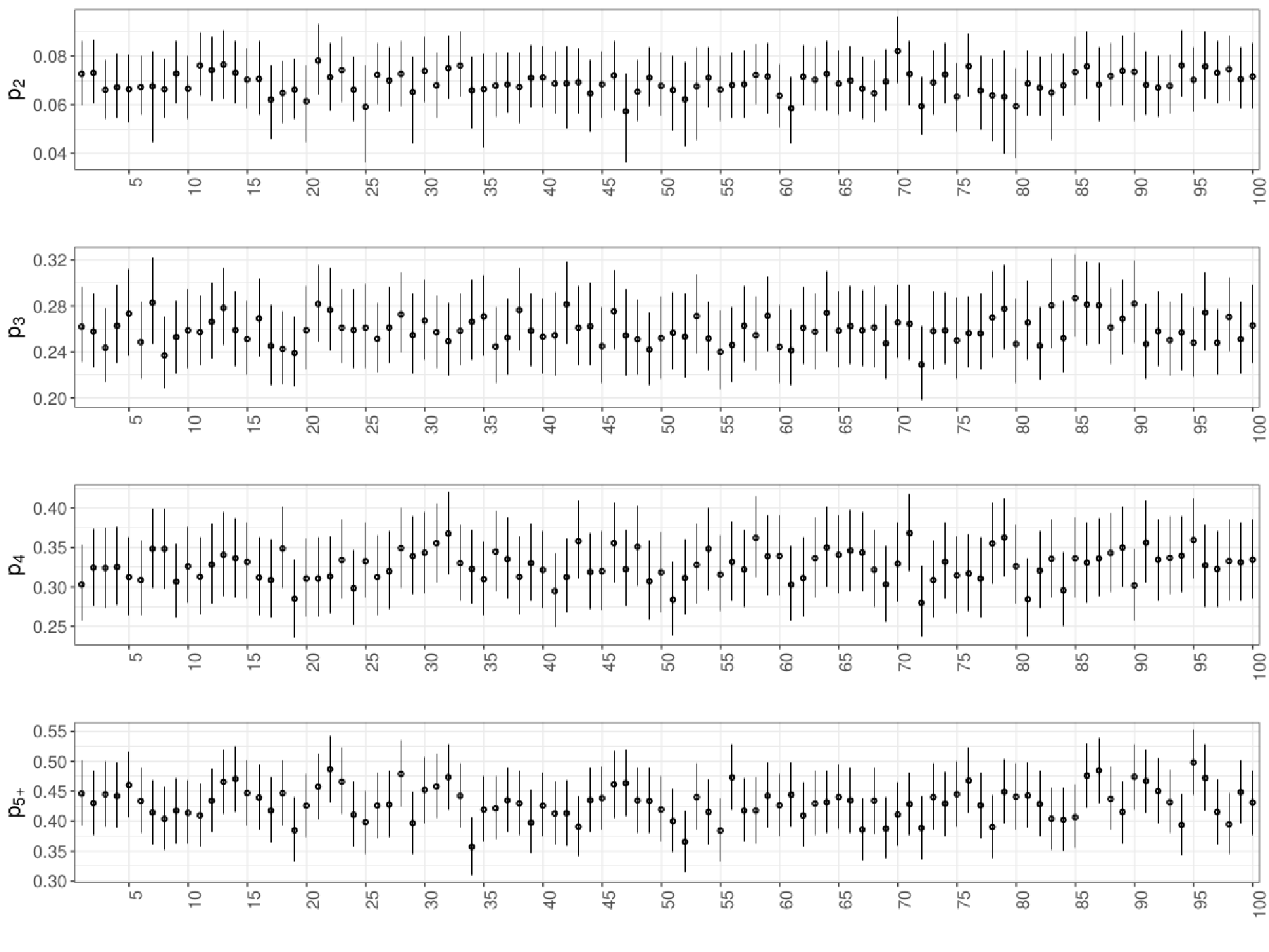}
\caption{\label{recap}Case study: Subposterior means and 95\% symmetric CIs for recapture probabilities ($p_a$) for each subsample for ages $a=2, \ldots, 5+$.}

\end{figure}

\begin{figure}[H]
\includegraphics[width=1\textwidth]{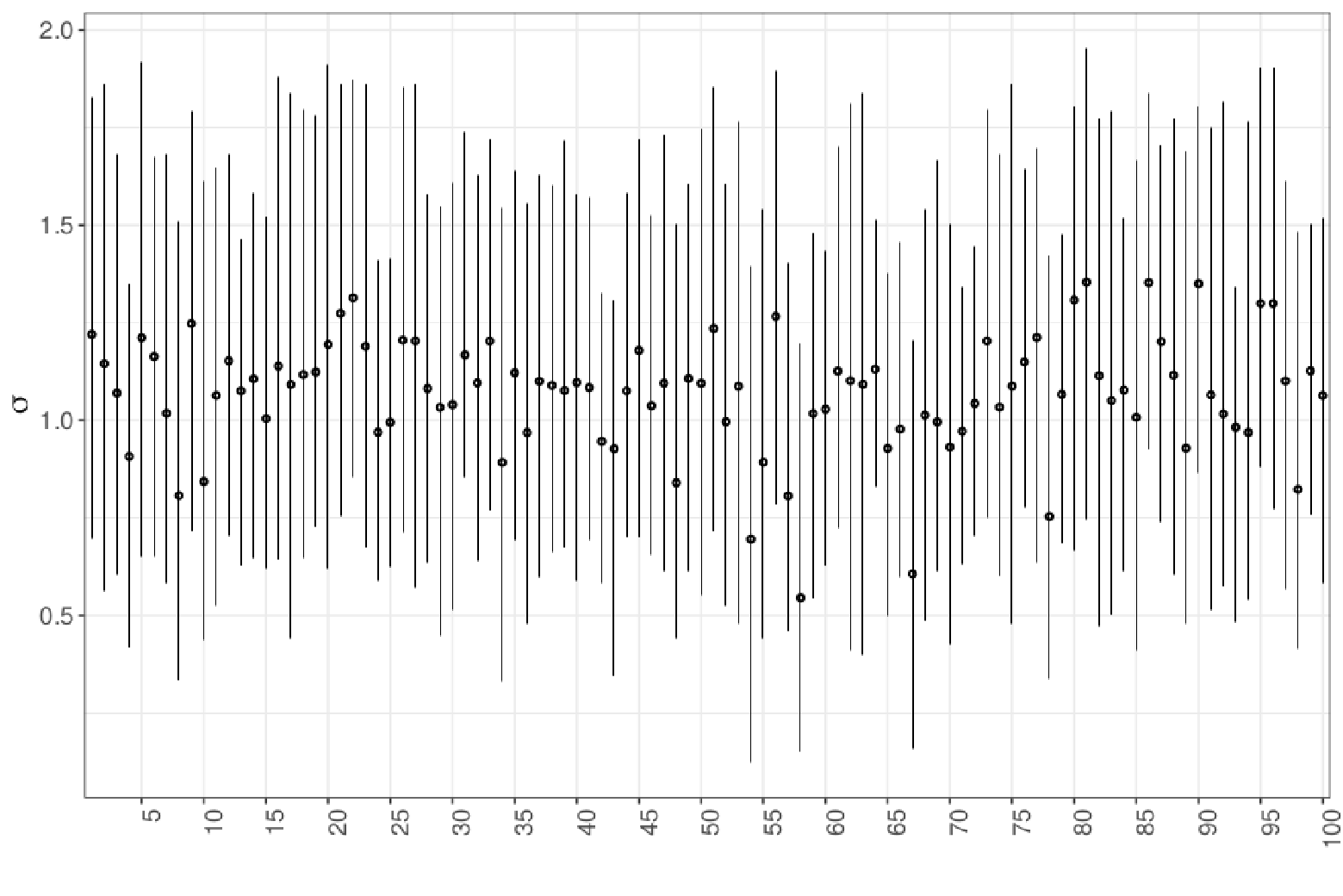}
\caption{\label{sigma}Case study: Subposterior means and 95\% symmetric CIs for the variance of the individual random effects ($\sigma$) for each subsample.}
\end{figure}

Figures \ref{alpha}-\ref{sigma} provide the estimated subposterior means and 95\% symmetric credible intervals (CIs) for the parameters across subsamples (i.e. the (uncorrected) posterior distribution of the parameters for the subsampled dataset, $\pi(\bftheta|\bfx^1)$) for the guillemot case study. 
Clearly, we can see that the marginal subposterior distributions for each of the parameters have a much greater variability than the posterior distribution for the full dataset, as expected. 



\section*{Web Appendix B}

\begin{figure}[H]
\includegraphics[width=1\textwidth]{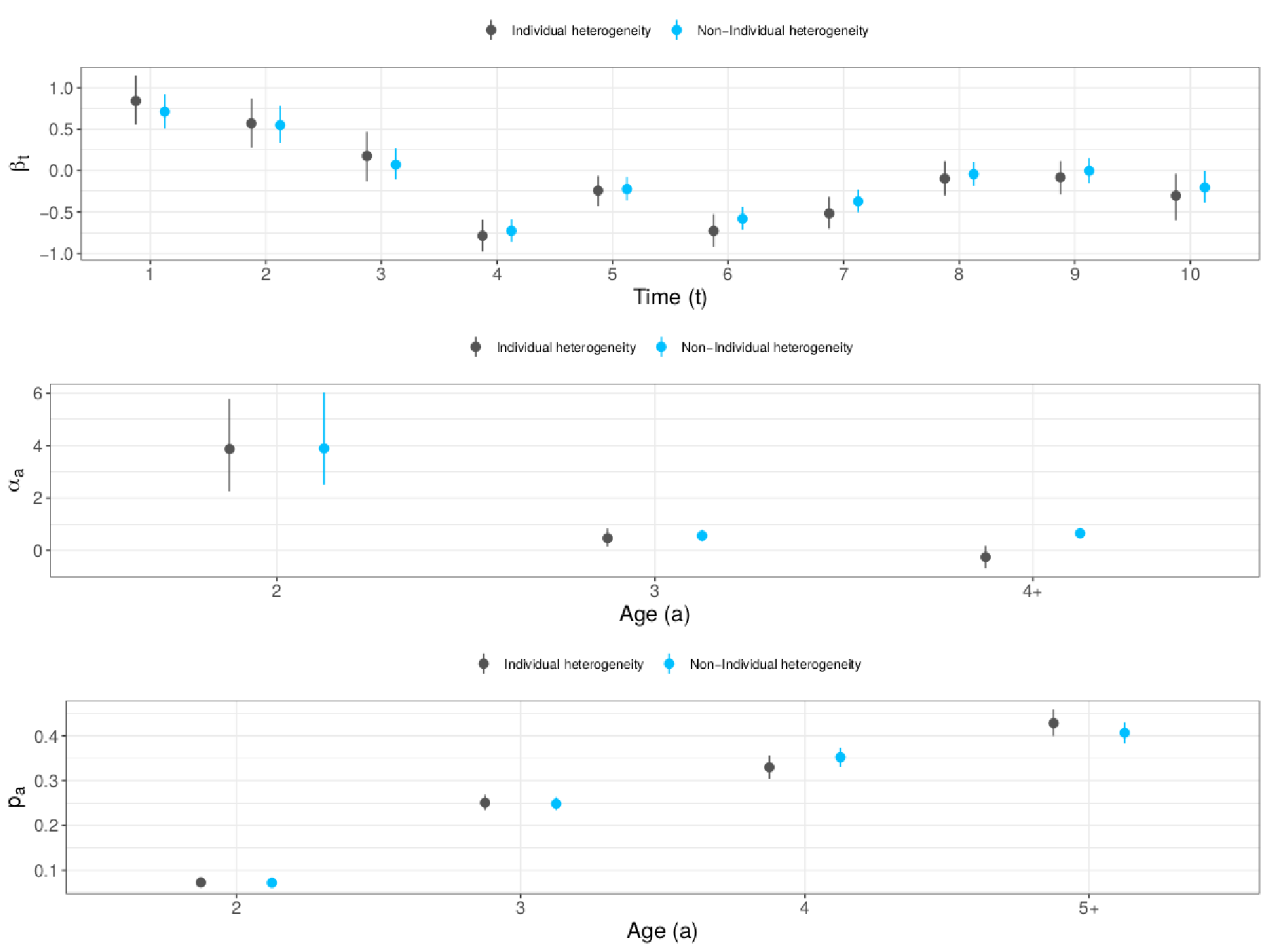}
\caption{\label{JABES_betas}Case study: Posterior means and 95\% CIs for model parameters for the model with individual heterogeneity (black) and previous model where individual heterogeneity was not considered (blue).}
\end{figure}
\begin{figure}[H]
\includegraphics[width=1\textwidth]{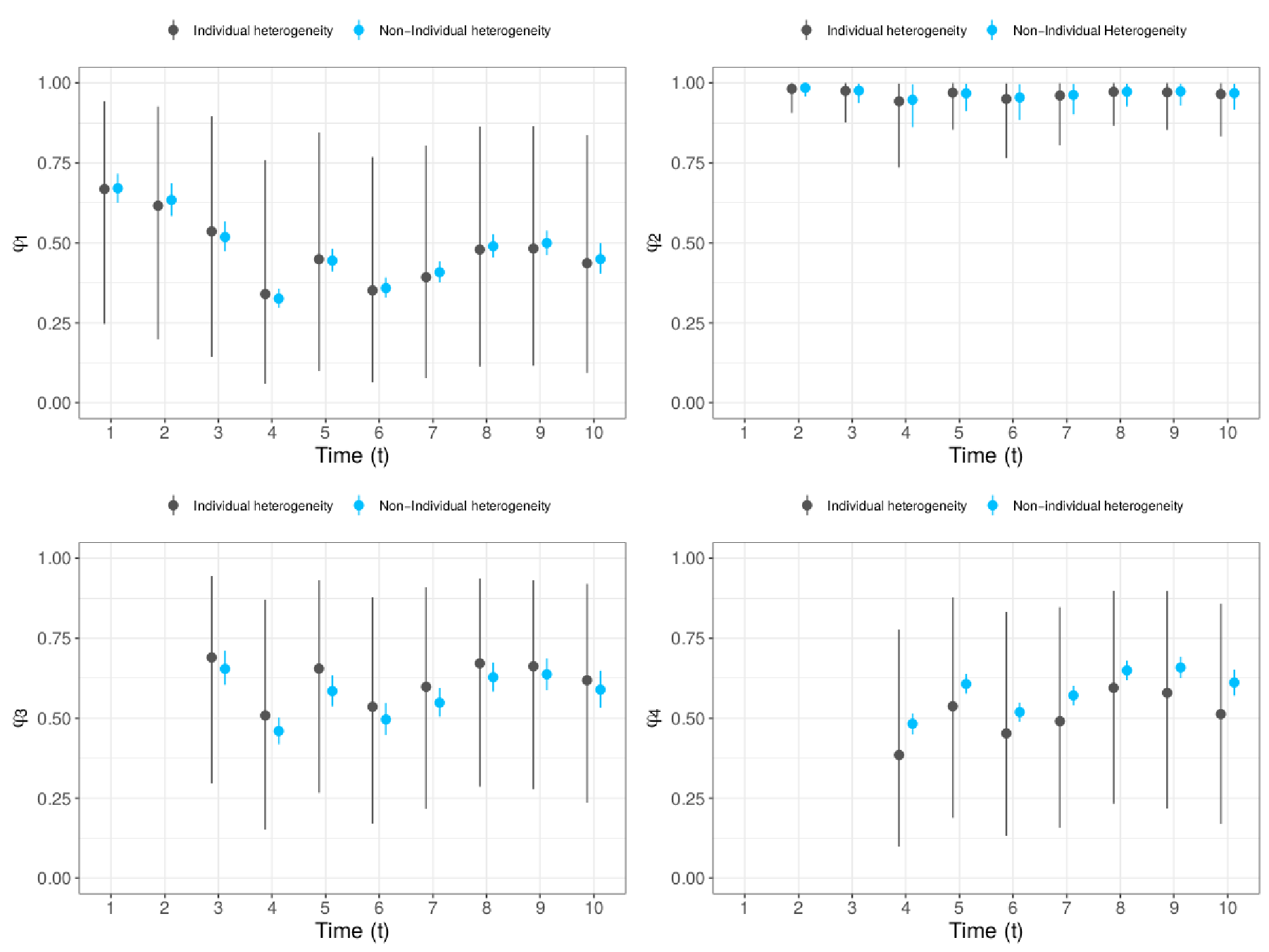}
\caption{\label{JABES_phis}Case study: Posterior means and 95\% CIs for age and time dependent survival probabilities for the model with individual heterogeneity (black) and previous model where individual heterogeneity was not considered (blue).}
\end{figure}

Figure \ref{JABES_betas} shows the posterior means and 95\% CIs for the model parameters for the individual heterogeneity model fitted within the paper, and the comparable model with the individual heterogeneity removed, as considered by \cite{Sarzo21}. The posterior estimates of the parameters are generally very similar, but with slightly larger CIs for the individual heterogeneity model. Figure \ref{JABES_phis} provides the associated age-dependent survival probabilities for each model. A relatively similar temporal dependence is observed over time, but with substantially wider CIs for the individual heterogeneity model. The wider CIs reflect the relatively large estimate of the (unobserved) individual heterogeneity component, with a posterior mean of the individual random effect standard deviation of 0.96 (on the logit scale), thus leading to a much wider posterior distribution for the survival probabilities for individuals. 

\bibliographystyle{chicago}
\bibliography{references.bib}